\documentclass{aa}  

\usepackage{amsmath}
\usepackage{amssymb}
\usepackage{graphicx}
\usepackage{color}
\usepackage{txfonts}

\usepackage{comment}

\usepackage{xcolor}
\definecolor{xlinkcolor}{cmyk}{1,1,0,0}
 \usepackage[bookmarks=false,         
     pdfnewwindow=true,      
     colorlinks=true,    
     linkcolor=xlinkcolor,     
     citecolor=xlinkcolor,     
     filecolor=xlinkcolor,  
     urlcolor=xlinkcolor,      
final=true
 ]{hyperref}

\usepackage{savesym}
\savesymbol{tablenum}
\usepackage{siunitx}
\usepackage{xspace}
\restoresymbol{SIX}{tablenum}
\usepackage{hyperref}
\usepackage{soul}

\newcommand\dbquote[1]{\textquotedblleft #1\textquotedblright}

\newcommand\partialdiff[1]{\frac{\partial}{\partial #1}}
\newcommand\stdiff[1]{\textrm{d} #1}

\newcommand\partiallogdiff[1]{\textrm{d}\mathrm{ln} #1/\textrm{d} \mathrm{ln}\, r}

\newcommand\Sigmagas{\Sigma_\textrm{g}}
\newcommand\Sigmadust{\Sigma_\textrm{d}}

\AtBeginDocument{}%
\AtBeginDocument{}%
\AtBeginDocument{}%
\AtBeginDocument{}%
\AtBeginDocument{}%

\begin{document} 
   \title{The external photoevaporation of structured protoplanetary disks}
   \titlerunning{External photoevaporation in structured disks}
   \author{Mat\'ias G\'arate\inst{1}          
        \and
          Paola Pinilla \inst{2}
        \and
          Thomas J. Haworth \inst{3}
        \and
          Stefano Facchini \inst{4}
          }
   \institute{
   $^{1}$Max-Planck-Institut f\"ur Astronomie, K\"onigstuhl 17, 69117, Heidelberg, Germany\\
   $^{2}$Mullard Space Science Laboratory, University College London, Holmbury St Mary, Dorking, Surrey RH5 6NT, UK\\
   $^{3}$Astronomy Unit, School of Physics and Astronomy, Queen Mary University of London, London E1 4NS, UK\\
   $^{4}$Universit\'a degli Studi di Milano, via Giovanni Celoria 16, 20133 Milano, Italy\\
  \email{garate@mpia.de, p.pinilla@ucl.ac.uk}
             }
   \date{}

  \abstract
   {The dust in planet-forming disks is known to evolve rapidly through growth and radial drift. In the high irradiation environments of massive star-forming regions where most stars form, external photoevaporation also contributes to rapid dispersal of disks. This raises the question of why we still observe quite high disk dust masses in massive star-forming regions.
   }
   {We test whether the presence of substructures is enough to explain the survival of the dust component and observed millimeter continuum emission in protoplanetary disks located within massive star-forming regions. We also characterize the dust content removed by the photoevaporative winds.
   }
   {
   We performed hydrodynamical simulations (including gas and dust evolution) of protoplanetary disks subject to irradiation fields of $F_{UV} = 10^2$, $10^3$, and $10^4\, G_0$, with different dust trap configurations.
   We used the FRIED grid to derive the mass loss rate for each irradiation field and disk properties, and then proceed to measure the evolution of the dust mass over time. For each simulation we estimated the continuum emission at $\lambda = 1.3\, \textrm{mm}$ along with the radii encompassing $90\%$ of the continuum flux, and characterized the dust size distribution entrained in the photoevaporative winds, in addition to the resulting far-ultraviolet (FUV) cross section.
   }
   {
   Our simulations show that the presence of dust traps can extend the lifetime of the dust component of the disk to a few millionyears if the FUV irradiation is $F_{UV} \lesssim 10^3 G_0$, but only if the dust traps are located inside the photoevaporative truncation radius. The dust component of a disk will be quickly dispersed if the FUV irradiation is strong ($10^4\, G_0$) or if the substructures are located outside the photoevaporation radius. We do find however, that the dust grains entrained with the photoevaporative winds may result in an absorption FUV cross section of $\sigma \approx 10^{-22}\, \textrm{cm}^2$ at early times of evolution ($<$0.1\,Myr), which is enough to trigger a self-shielding effect that reduces the total mass loss rate, and slow down the disk dispersal in a negative feedback loop process.
   }
   {}

   \keywords{accretion, accretion disks -- 
            protoplanetary disks --
            hydrodynamics --  
            methods: numerical}

   \maketitle
%

\section{Introduction} \label{sec_Intro}
%
Protoplanetary disks are composed of the gas and dust material that orbits around newly formed stars. 
The evolution of isolated protoplanetary disks is often modeled by accounting for internal processes such as viscous accretion driven by magneto-rotational instabilities (MRI), magneto-hydrodynamic (MHD) winds, and/or thermal photoevaporative winds due to the irradiation from the central star \citep{Lynden-Bell1974, Balbus1991, Blandford1982, Clarke2001, Pascucci2023_review}.\par
However, growing observational evidence suggest that the irradiation from the environment also plays a significant role in the disk evolution in dense stellar regions such as the Orion Nebular Cluster, Cygnus, and Upper Sco \citep[see][Anania et al., in prep.]{Guarcello2016, Otter2021, vanTerwisga2023}. In these regions, observations report particularly compact disk sizes, which points towards photoevaporative truncation, and disks that exhibit strong morphological signatures of mass loss \citep[e.g.][]{Odell1993, McCullough1995, Bally1998, mann2014, ballering2023}.\par

The observational evidence is also consistent with theoretical models that predict significant mass loss rates due to the environmental radiation in a process known as external photoevaporation. \citep[e.g.][]{Scally2001, Adams2004, Anderson2013, Facchini2016, Haworth2018}. However, one of the open questions regarding the disks in these highly irradiated regions is why these disks have not yet dispersed if the mass loss rates experienced are so high, which has also been dubbed as the \dbquote{proplyd lifetime problem} \citep{HenneyOdell1999}.\par

Some of the solutions proposed to this problem include taking into account different star formation events in a given region, which means that some of the disks are actually younger than the cluster itself, and also considering that stars (along with their protoplanetary disks) migrate within the cluster, and therefore they experience a varying ultra-violet (UV) irradiation during their lifetime \citep{Winter2019b_Cyg, Winter2019}. Additionally, disks may be shielded from external irradiation by an optically thick envelope during the early stages of evolution, delaying the starting time of the photoevaporation process \citep{Qiao2022, Wilhelm2023}.\par

However, besides the survival of the gas component which is directly dispersed by external photoevaporation, it is also necessary to explain the survival of the dust component, which is observed in the millimeter continuum \citep{Eisner2018, Otter2021}. 
Numerical models by \cite{Sellek2020} have shown that drift is even more efficient in disks truncated by external photoevaporation, with typical depletion timescales of $t_\textrm{depletion} \approx \SI{2e5}{yr}$ (defined as the timescale in which $99\%$ of the initial dust mass is lost).\par
In order to explain the detected millimeter fluxes and sizes, it is necessary to retain the dust for longer timescales, with dust-trapping substructures for example \citep{Whipple1972, Pinilla2012}. Except for very young class 0/I objects \citep[see the recent eDisk sample]{Ohashi2023_eDisk}, substructures such as rings and gaps appear to be a common feature in protoplanetary disks \citep[e.g. the DSHARP sample][]{Andrews2018}, occurring even in compact disks \citep[][Miley et al., in prep.]{Zhang2023}.
Despite this, there are currently no models that study the evolution and observability of a sub-structured disk subject to external photoevaporation.
In particular, it is not clear if the solid material that is concentrated at dust traps will be able to survive the photoevaporative dispersal, or if it will be dragged along with the gas component in the thermal winds.

The goal of this paper is to characterize the evolution of the dust size distribution, the flux in the millimeter continuum, and the estimated disk size. We performed numerical simulations of disks that are subject to external photoevaporation, in cases with and without gap-(ring-)like structures.

We are interested on whether the dust traps located at the edge of gap-like structures can survive the photoevaporative dispersal process, for different UV fluxes and for different trap locations. 
Though we expect only the small grains to be entrained in the photoevaporative wind, the fragmentation of larger dust and uncertain growth timescale of small dust makes the outcome unclear. We also test the survival of such dust traps in the particular case of an increasing UV flux, motivated by the model \citet{Winter2019} in which protoplanetary disks experience a variable UV irradiation during their lifetime. Finally, from our simulations we also track the distribution of the dust grains entrained with the winds, and use that information to re-estimate the effective opacities at UV wavelengths, which could result in a self-shielding effect that would protect the disk from the environmental irradiation \citep{Facchini2016, Qiao2022}.

The paper is structured as follows:
In \autoref{sec_Theory} we describe our gas and dust evolution model, which includes the growth and fragmentation of multiple grain species, the prescription for the gap-like structures, and the external photoevaporation model using the FRIED mass loss rate grid \citep{Haworth2018, Sellek2020}. 
In \autoref{sec_Setup} we describe the simulations, the parameter space, and the calculation of different observable quantities.
\autoref{sec_Results} shows our results, including a review of the difference between photoevaporating and non-photoevaporating disks, the evolution of our simulations for the different parameters, and the description of the dust content in the wind.
\autoref{sec_Discussion} discusses the implications of our results and how they compare to observations of highly irradiated regions.

\section{Disk model} \label{sec_Theory}
We used the code \texttt{DustPy}\footnote{ \href{https://github.com/stammler/DustPy}{github.com/stammler/DustPy}, version 0.5.6 was used for this work.} \citep{Stammler2022} to simulate the gas and dust evolution of protoplanetary disks, and include the effects of external photoevaporation. \par

\subsection{Gas evolution}

\begin{figure}
\centering
\includegraphics[width=90mm]{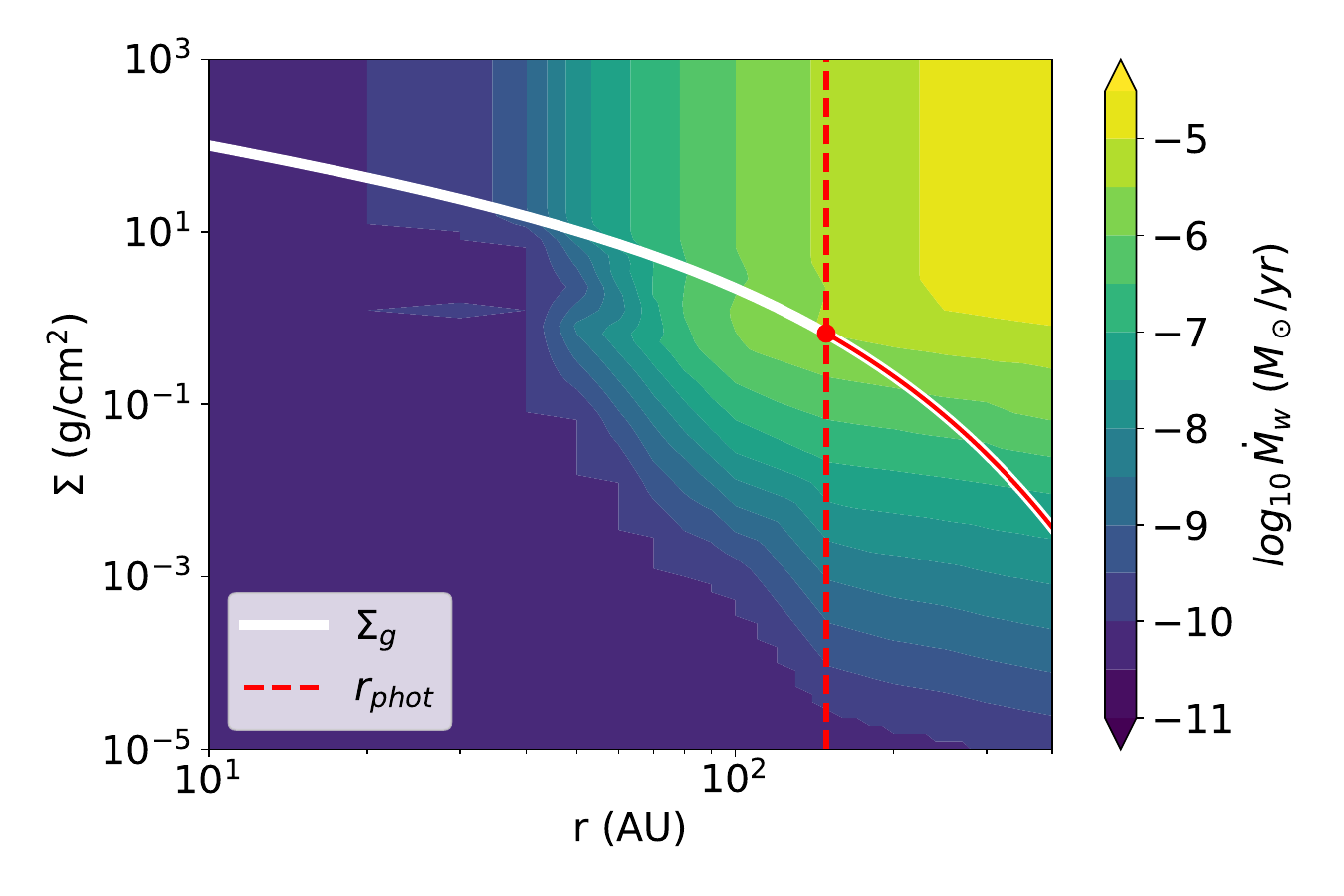}
 \caption{
 Example mass loss rate grid for a disk orbiting a $\SI{1}{M_\odot}$ star subject to an irradiation of $F_{UV} = 10^3\, G_0$. The grid determines the gas mass loss $\dot{M}_\textrm{w}$ as a function of the gas surface density $\Sigmagas$ and radii $r$, using the FRIED grid from \citet{Haworth2018} and the implementation of \citet{Sellek2020}. 
 The solid white line, shows an example gas surface density profile, the dashed vertical line indicates the photoevaporative truncation radius $r_{phot}$, and the solid red line indicates the regions which are subject to external photoevaporation. The photoevaporation radius evolves with the simulation, and corresponds to the maximum of the mass loss rate along the current surface density profile.}
 \label{Fig_Demo_MassLossMap}
\end{figure}

The gas component evolves through viscous spreading and mass loss by external photoevaporation, through the following equation:

\begin{equation} \label{eq_gas_advection}
    \partialdiff{t}\Sigma_{\textrm{g}} + \frac{1}{r} \partialdiff{r} (r \, \Sigma_{\textrm{g}} \, \varv_\nu)= -\dot{\Sigma}_\textrm{g,w},
\end{equation}
where $r$ is the radial distance to the star, $\Sigmagas$ is the gas surface density, and $\varv_\nu$ is the radial viscous velocity \citep{Pringle1981}:
\begin{equation} \label{eq_visc_velocity}
\varv_\nu = -\frac{3}{\Sigmagas \sqrt{r}}\partialdiff{r}(\nu_\alpha \, \Sigmagas \, \sqrt{r}),
\end{equation}
with $\nu_\alpha$ the kinematic viscosity:
\begin{equation}
 \nu_\alpha = \alpha c_s h_\textrm{g},  
\end{equation}
which depends on the sound speed $c_s$, the pressure scale height $h_\textrm{g}$, and the turbulence parameter $\alpha$ \citep{Shakura1973}.
The adiabatic sound speed is defined as:
\begin{equation}
    c_s = \sqrt{\gamma \frac{k_B T}{\mu m_p}},
\end{equation}
where $\gamma$ is the adiabatic index, $T$ is the gas temperature, $k_B$ is the Boltzmann constant, $\mu$ is the mean molecular weight, and $m_p$ is the proton mass.\par
The surface density loss rate $\dot{\Sigma}_\textrm{g,w}$, was derived from the \texttt{FRIED} grid from \cite{Haworth2018}, following the implementation of \citet[][see their equations 5 - 7]{Sellek2020}. 
To describe it briefly, this method first predicts the expected total mass loss rate $\dot{M}_\textrm{w}$, for a disk with surface density $\Sigmagas$, around a star with mass $M_*$, subject to an external FUV radiation field $F_\textrm{UV}$, along with the photoevaporative radius $r_\textrm{phot}$ which is located at the interface between the optically thick and optically thin regions in the outer disk. 
Then, the total mass loss rate $\dot{M}_\textrm{w}$ is distributed across all the grid cells in the photoevaporative region ($r \geq r_\textrm{phot}$), according to the material available in each grid cell. \autoref{Fig_Demo_MassLossMap} shows an example of the mass loss grid, with a surface density profile overlaid on top to illustrate the location of the photoevaporation radius and the regions affected by the mass loss.\par
This implementation assumes that the mass loss by external photoevaporation occurs only in the outer regions of the disk, and loss from the inner regions is negligible. For more details about the implementation, we refer to their original paper.\footnote{The implementation of external photoevaporation in \texttt{DustPy} is available at:\\ \href{https://github.com/matgarate/dustpy_module_externalPhotoevaporation}{github.com/matgarate/dustpy\textunderscore module\textunderscore externalPhotoevaporation},\\ and will also be included within the \texttt{DustPy} library package at:\\ \href{https://github.com/stammler/dustpylib}{github.com/stammler/dustpylib}.} For simplicity, we did not include internal MHD or photoevaporative winds, which would be launched from smaller radii. \\

\subsection{Dust dynamics}

\texttt{DustPy} tracks the evolution of multiple grain sizes, that can grow through sticking, fragmentation into smaller species, drift towards the local pressure maximum, and diffuse according to the concentration gradient, following the model from \citet{Birnstiel2010}.
The corresponding advection-diffusion equation is:
\begin{equation} \label{eq_dust_advection}
    \partialdiff{t}  \Sigma_{\textrm{d}} + \frac{1}{r}\partialdiff{r} (r \, \Sigma_{\textrm{d}} \, \varv_{\textrm{d}}) - \partialdiff{r} \left(r D_\textrm{d} \Sigmagas \partialdiff{r}\left(\frac{\Sigmadust}{\Sigmagas}\right)\right)= -\dot{\Sigma}_\textrm{d,w},
\end{equation}
where $\Sigmadust$ is the dust surface density, $\varv_\textrm{d}$ is the corresponding radial velocity, $D_d$ is the dust diffusivity, and $\dot{\Sigma}_\textrm{d, w}$ is the dust loss by wind entrainment. We note that we solve this equation for every dust size bin (along with the coagulation equation, see \autoref{Sec_Model_GrainGrowth}) and therefore all dust quantities are dependent on the grain size.\par

Overall, the dust dynamics can be described by its particle size $a$, or more specifically, by the Stokes number $\mathrm{St}$, which measures the coupling between gas and dust with:
\begin{equation} \label{eq_StokesMidplane}
    \textrm{St} = \frac{\pi}{2}\frac{a\, \rho_s}{\Sigma_g} \cdot
    \begin{cases}
				1 & \lambda_\textrm{mfp}/a \geq 4/9\, -\, \mathrm{Epstein},\\
                \frac{4}{9} \frac{a}{\lambda_\textrm{mfp}} & \lambda_\textrm{mfp}/a < 4/9\, -\,  \mathrm{Stokes\, I},
	\end{cases} 
\end{equation}
with $\rho_s$ the material density of the dust, and $\lambda_\textrm{mfp}$ the mean free path of the gas molecules, where the latter is used to determine the corresponding drag regime (Epstein or Stokes I).\par

Given the Stokes number, the radial advection velocity of the dust is defined by:
\begin{equation} \label{eq_dust_radial_velocity}
    \varv_\textrm{d} = \frac{1}{1 + \mathrm{St}^2} \varv_\nu -  \frac{2 \mathrm{St}}{1 + \mathrm{St}^2} \eta \varv_k.
\end{equation}
Which means that small grains ($\mathrm{St} \ll 1$) become coupled to the gas motion $\varv_\nu$, large boulders ($\mathrm{St} \gg 1$) become decoupled from the gas, and mid-sized pebbles ($\mathrm{St}\sim1$) drift towards the pressure maximum with a velocity of $\varv_\textrm{d} \approx -\eta \varv_K$, where $\eta = -\, (1/2)\,  (h_\textrm{g} / r)^2\, \partiallogdiff{P}$ measures the relative difference between the gas orbital speed and the local Keplerian speed $\varv_K$ due to the pressure support of the gas $P = \Sigmagas c_s^2/(\sqrt{2\pi} h_\textrm{g})$ \citep{Weidenschilling1977, Nakagawa1986, Takeuchi2002}. \par
We note that the effect of the pressure gradient on the dust dynamics is measured at the midplane, since large particles settle down to smaller scale heights than the gas, with $h_\textrm{d} = h_\textrm{g} \sqrt{\alpha/(\alpha + \mathrm{St})}$ \citep{Dubrelle1995}. Observational evidence of effective dust settling  has recently been confirmed by observations of a handful of edge-on disks at different wavelengths\citep{Villenave2020}. The radial diffusivity is characterized by $D_\textrm{d} = \nu_\alpha/(\mathrm{St^2} + 1)$, where smaller particles diffuse more easily than larger ones \citep{Youdin2007}.\par

To include the effect of dust entrainment in the photoevaporative wind, we followed the prescription from \citet{Sellek2020}, where grains smaller than the entrainment size can be lost with the wind:
\begin{equation} \label{eq_entrainment_size}
    a_\textrm{ent} = \sqrt{\frac{8}{\pi}}\frac{c_s}{G M_*} \frac{\dot{M}_\textrm{w}}{\Omega_* \rho_s},
\end{equation}
where $\Omega_* = 4\pi h_\textrm{g}/\sqrt{h_\textrm{g}^2 + r^2}$ is the solid angle covered by the disk outer edge, as seen from the star.
Then, the surface density loss rate for dust grains with sizes $a \leq a_\textrm{ent}$ is:
\begin{equation} \label{eq_dust_entrainment}
   \dot{\Sigma}_\textrm{d,w}(a) = \epsilon(a)  \dot{\Sigma}_\textrm{g,w},
\end{equation}
with $\epsilon(a)$ the dust-to-gas ratio of the grains in the bin size $a$.\par

In terms of disk evolution, this could lead to an outcome where the dust grains in the outer regions are either entrained at early phases during the disk lifetime, or grow and drift before they can be entrained with the wind. The study of \citet{Sellek2020} suggests that the former scenario occurs in protoplanetary disks, though differences may arise when including the full coagulation of multiple grain sizes in the model, instead of the two-population approximation from \citet{Birnstiel2012}.

\subsection{Substructures and  dust traps}
To include the effects of substructures capable of trapping dust grains (e.g. such as the gaps formed by giant planets), we used the same approach as in \cite{Stadler2022}, which implements a perturbation in the viscous $\alpha$ profile, that in turn creates a gap-like structure in the gas surface density profile.
The perturbation in the turbulence has the shape of a Gaussian bump:
\begin{equation} \label{eq_alpha_bump}
    \alpha(r) = \alpha_0 \times \left(1 + A_\textrm{gap} \exp\left(-\frac{\left(r - r_\textrm{gap}\right)^2}{2w_\textrm{gap}^2}   \right)\right),
\end{equation}
where, $\alpha_0$ is the turbulence base value, and $A_\textrm{gap}$, $r_\textrm{gap}$, and $w_\textrm{gap}$ are the gap amplitude, location, and width, respectively. From viscous evolution theory,  a power-law gas surface density profile is scaled approximately by a factor of $\Sigmagas (r) \propto \alpha_0 / \alpha(r)$, assuming that the disk is in steady state accretion.\par
The presence of a gap-like substructure then leads to the formation of a pressure maximum, where large particles ($\mathrm{St} \gtrsim \alpha$) can be easily trapped \citep[][see also see \autoref{eq_dust_radial_velocity}]{Pinilla2012}, though we note that small particles would still be able to pass through the gap by coupling with the gas component.\par

\subsection{Grain growth} \label{Sec_Model_GrainGrowth}
Dust growth was computed by solving the coagulation equation \citep{Smoluchowski1916}, that accounts for the result of the collision between two grain species given their relative velocities and sizes, which can be the sticking between particles, the fragmentation of both species, and the erosion in case of a significant size difference \citep[for more details we refer to][]{Birnstiel2010, Stammler2022}.\par

There are two characteristic regimes of grain growth: the drift-limited, when particles drift inward faster than they can grow, and the fragmentation-limited, when particles collide at speeds higher than the fragmentation threshold of the material $\varv_\textrm{frag}$ \citep{Brauer2008, Birnstiel2009, Birnstiel2012}. In particular, in pressure maxima, such as the one in the outer edge of a gap, the contribution of drift to both the radial motion and the relative collision velocities between dust grains is cancelled, allowing the particles to locally accumulate and grow into larger sizes until they reach the size limit given by the fragmentation barrier.\par

\section{Simulation setup} \label{sec_Setup}

In this section we describe the initial conditions of our simulations, the parameter space explored, and the post-processing method used to derive the observable fluxes from dust grain size distribution.

\subsection{Initial conditions and grid resolution}
The initial surface density profile was set according to a modified version of the \citet{Lynden-Bell1974} self-similar solution, that include gap-like substructures that are consistent with the turbulence radial profile (see \autoref{eq_alpha_bump}):
\begin{equation} \label{eq_LBPprofile}
    \Sigmagas(r) = \frac{M_\textrm{disk}}{2\pi r_c^2} \left(\frac{r}{r_c}\right)^{-1} \exp(-r/r_c) \frac{\alpha_0}{\alpha(r)},
\end{equation}
where $M_\textrm{disk}$ and $r_c$ are the initial disk mass and characteristic radius.\par
The gas temperature for the simulations follows a power law profile, with:
\begin{equation} \label{eq_temperature_profile}
    T_\textrm{g}(r) = T_0 \left(\frac{r}{\SI{1}{AU}}\right)^{-1/2},
\end{equation}
where $T_0$ is the temperature at \SI{1}{AU}. We note that this profile assumes that the heating is dominated by the passive stellar irradiation on the disk, and neglects the contribution of accretion heating, and the radiation from nearby stars.The dust distribution is initialized assuming a uniform constant dust-to-gas ratio, with $\Sigmadust = \epsilon_0 \Sigmagas$, following the ISM size distribution from \citep{Mathis1977}, which goes from $\SI{0.5}{\mu m}$ to an initial maximum grain size of $a_0 = \SI{1}{\mu m}$.\par
The radial grid was set to be linearly spaced with $r^{1/2}$, and going from \SI{2.5}{AU} to \SI{500}{AU}, with $n_r = 200$ grid cells. The FRIED grid is tabulated out to \SI{400}{AU}, however, in our models the photoevaporative truncation radius is always smaller than the outer radial grid boundary of the FRIED grid and no extrapolation in the 400-500 AU range is needed. The grid for the dust size distribution was set with a logarithmic spacing, going from approx. \SI{0.5}{\mu m} to \SI{50}{cm} with $n_m = 127$ grid cells, such there are 7 grid cells for each order of magnitude in mass in order to reliably solve the grain coagulation \citep{Ohtsuki1990, Drazkowska2014, Stammler2022}. We evolved the simulations from time $t = 0$ up to $\SI{5}{Myr}$, though some disks may fully disperse earlier due to the influence of external photoevaporation.

\subsection{Parameter space}
In this work, our main focus is to explore on how the gap presence and its location affects the retention of solids in protoplanetary disks that are subject to different FUV radiation environments.For the fiducial model we took a \SI{1}{M_\odot} mass star, surrounded by a disk with mass of $M_\textrm{disk} = 0.1 M_*$, and an initial characteristic radius of $r_c = \SI{90}{AU}$. We started our study with an initial comparison against disks without photoevaporation, to have an overview of the key aspects of the disk evolution. 

For the main parameter space exploration, gap-like substructures can be located at $1/3\, r_c$ (inner trap) or $2/3\, r_c$ (outer trap), or completely absent. ($A_\textrm{gap} = 0$, no traps). With $r_c = \SI{90}{AU}$, this means that the gaps are located either at $\SI{30}{AU}$ or $\SI{60}{AU}$ in our simulations. \cite{Bae2022}  compiled data of disks with observed substructures, showing that most of the rings have been found between 20-60\,AU (their histogram in Fig. 3d), as assumed in this work. Some disks have substructures up to 100-200\,AU, which are outside the  photoevaporation radius in our models. The disk can be subject to external photoevaporation due to far ultra-violet (FUV) fluxes of $10^2\, G_0$\footnote{The FUV radiation field strength is usually measured using the Habing unit,  defined as $1$G$_0=1.6\times10^{-3}$ erg\,s$^{-1}$\,cm$^{-2}$ integrated over the wavelength range 912--2400\AA. 1\,G$_0$ is representative of the mean interstellar FUV radiation field in the Solar neighbourhood. } (weak), $10^3\, G_0$ (medium), or $10^4\, G_0$ (strong) \citep[where $G_0$ corresponds to the local interstellar radiation field ][]{Habing1968}. We note that in low mass star-forming regions such as Taurus/Lupus the external FUV radiation field is of order 1-100\,G$_0$. In more massive stellar clusters such as the Orion Nebular Cluster the FUV radiation field strength that disks are exposed to ranges from $1-10^7$\,G$_0$. Previous work has suggested that the most common experienced UV environment is around $10^3$\,G$_0$ 
 \citep{2008ApJ...675.1361F, 2020MNRAS.491..903W}. 
Additionally, we  also study how the disk evolution changes for a star with lower mass (correspondingly with lower disk temperature), a disk that is initially more compact, and a disk that is subject to a FUV radiation field that increases with time \citep[in order to mimic, for example the effect of migration within a stellar cluster, or the clearing of the original molecular cloud][]{Winter2019, Qiao2022}. The complete parameter space and the additional disk physical parameters can be found in \autoref{Table_OverviewParameters}.\par

\begin{table}[h]
\caption{Parameter space. Fiducial values are highlighted in boldface.}
\label{Table_OverviewParameters}
\begin{tabular}{l l c l}
 \hline \hline
Symbol & Description & Value & Unit \\
 \hline
$F_\textrm{UV}$ & External FUV Flux & $10^2$, $\mathbf{10^3}$, $10^4$  & $G_0$ \\
$M_*$ & Stellar mass & 0.3, \textbf{1.0} & $M_\odot$ \\
$M_\mathrm{disk}$ & Initial disk mass & 0.1 & $M_*$ \\
$r_\mathrm{c}$ & Characteristic radius & 45, \textbf{90} & AU \\
$r_\mathrm{gap}$ & Gap locations & \textbf{1/3}, 2/3 & $r_\mathrm{c}$ \\
$A_\mathrm{gap}$ & Trap amplitude & \textbf{0}, \textbf{4} & - \\
$T_0$ & Temperature at \SI{1}{AU} & 125, \textbf{205} & K \\
$\alpha_0$ & Viscosity parameter &  $10^{-3}$ & - \\
$\gamma$ & Adiabatic index & 1.4 & - \\
$\mu$ & Mean molecular weight & 2.3 & - \\
$\epsilon_0$ & Initial dust-to-gas ratio & $0.01$ & - \\
$a_0$ & ISM maximum grain size  & 1 & $\mu$m \\
$\varv_\textrm{frag}$ & Fragmentation velocity & 10 & m\,s$^{-1}$ \\
$\rho_s$ & Grain material density & 1.67 & g\,cm$^{-3}$ \\
\hline
\end{tabular}
\vspace{-3.0mm}
\end{table}

\subsection{Fluxes in the millimeter continuum}
To compare our models with observations, we obtain the grain size distribution from the dust evolution simulations with \texttt{DustPy}, which is then used to compute the intensity profile and the total flux in the millimeter continuum, at $\lambda = \SI{1.3}{mm}$. We assume that the dust grains follow the opacity model from \citet{Ricci2010}, and are composed of water ice, carbon and silicates \citep{Zubko1996, Draine2003, warren&brandt08}.\par
With the opacity profile $k_\nu(a)$, with $\nu$ the frequency, we can calculate the optical depth at every radius as:
\begin{equation}
    \tau_\nu = \sum_a \kappa_\nu(a)\, \Sigmadust(a),
\end{equation} 
the intensity profile with:
\begin{equation} \label{eq_Flux_OpticallyThin}
    I_\nu = B_\nu(T) \left(1 - \exp(-\tau_\nu) \right), 
\end{equation}
and the total flux as $F_\nu = \int I_\nu\, \stdiff{\Omega}$, where $\stdiff{\Omega}$ is the differential of the solid angle covered by the disk in the sky, and $B_\nu(T)$ is the emission of a black body with temperature $T$. 
For the purpose of this work, we assumed our disks to be at a distance of \SI{400}{pc}, which is the approximate distance to the Orion Nebular Cluster \citep[ONC,][]{Hirota2007, Kounkel2017}, and to account for observational limitations, we also assume a beam size of $\SI{40}{mas}$ and a sensitivity threshold of \SI{0.1}{mJy\, beam^{-1}} (this is used to calculate the measured disk size $r_{90\%}$, and the intensity profile $I_\nu$).

While for the gas and dust evolution, both of them have the same temperature (Eq.~\ref{eq_temperature_profile}),  for the calculations of the millimetre fluxes, we assume that there is a background temperature of $T_\textrm{b} = \SI{20}{K}$, to account for the irradiation from massive stars in the interstellar space, thus: 

\begin{equation}
    T = \sqrt[4]{T_\textrm{g}^4 + T_\textrm{b}^4}.
\end{equation}

\section{Results} \label{sec_Results}
%
In this section we show how the mass evolution, dust size distribution, and observable properties such as the intensity profile, total flux and the observable dust disk size change due to the effect of external photoevaporation and substructures within our parameter space.\par

\subsection{Fiducial comparison} \label{sec_Resuts_Demo}

\begin{figure}
\centering
\includegraphics[width=90mm]{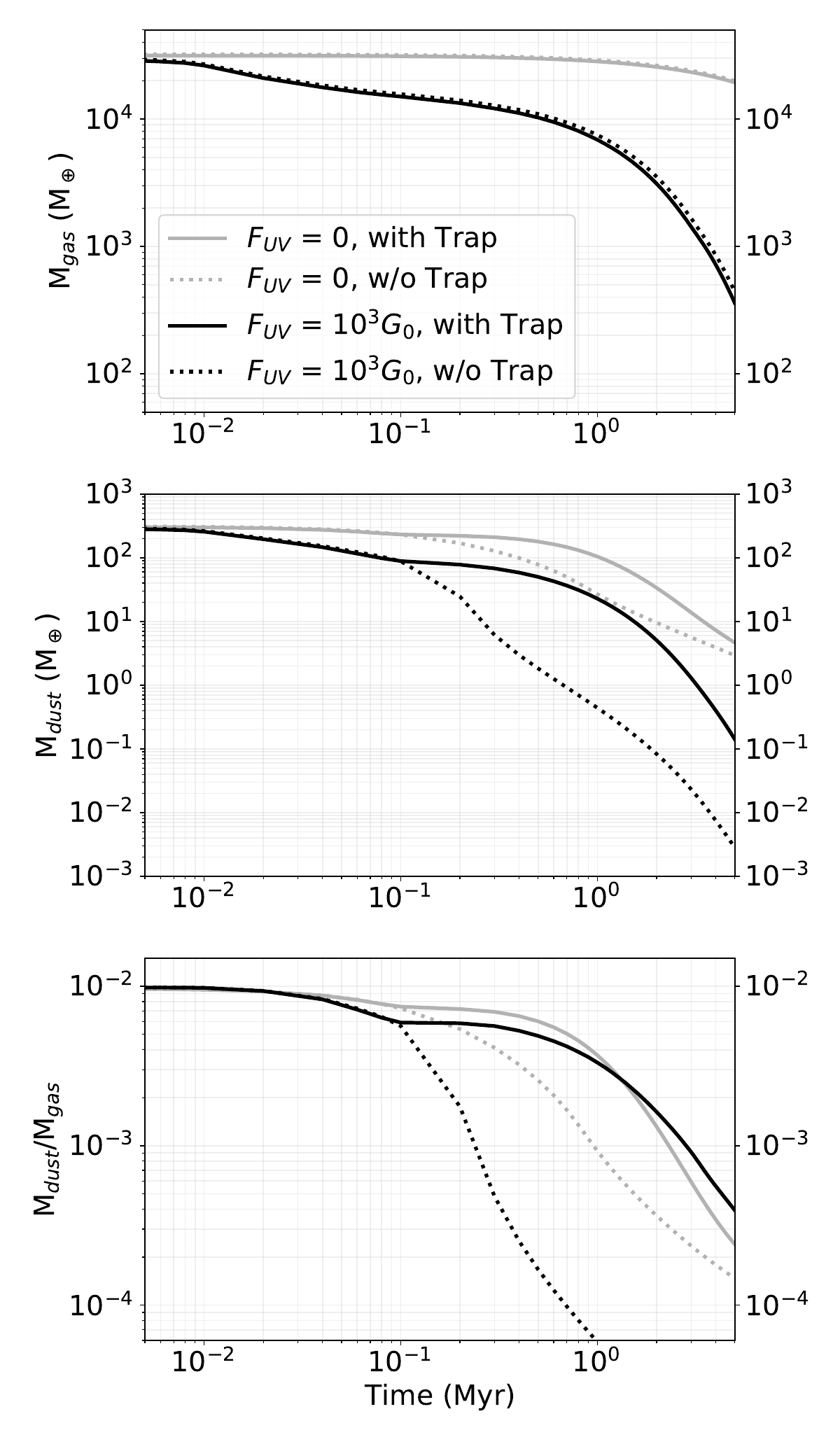}
 \caption{
 Gas and dust mass evolution (top and middle panels), and the global dust-to-gas ratio (bottom panel). The plot shows the evolution of a disk with and without the influence of FUV external photoevaporation (black vs. gray lines), for disks with and without a gap-like substructure (solid vs. dotted lines).
 For this comparison, the disk is around a $M_\odot$ mass star with an initial size of $r_c = \SI{90}{AU}$. The gap substructures are located at $r_\textrm{gap} = 1/3 r_c$, and the irradiation field is $F_\textrm{UV} = 10^3\, G_0$.
 }
 \label{Fig_Demo_MassEvolution}
\end{figure}

\begin{figure*}
\centering
\includegraphics[width=180mm]{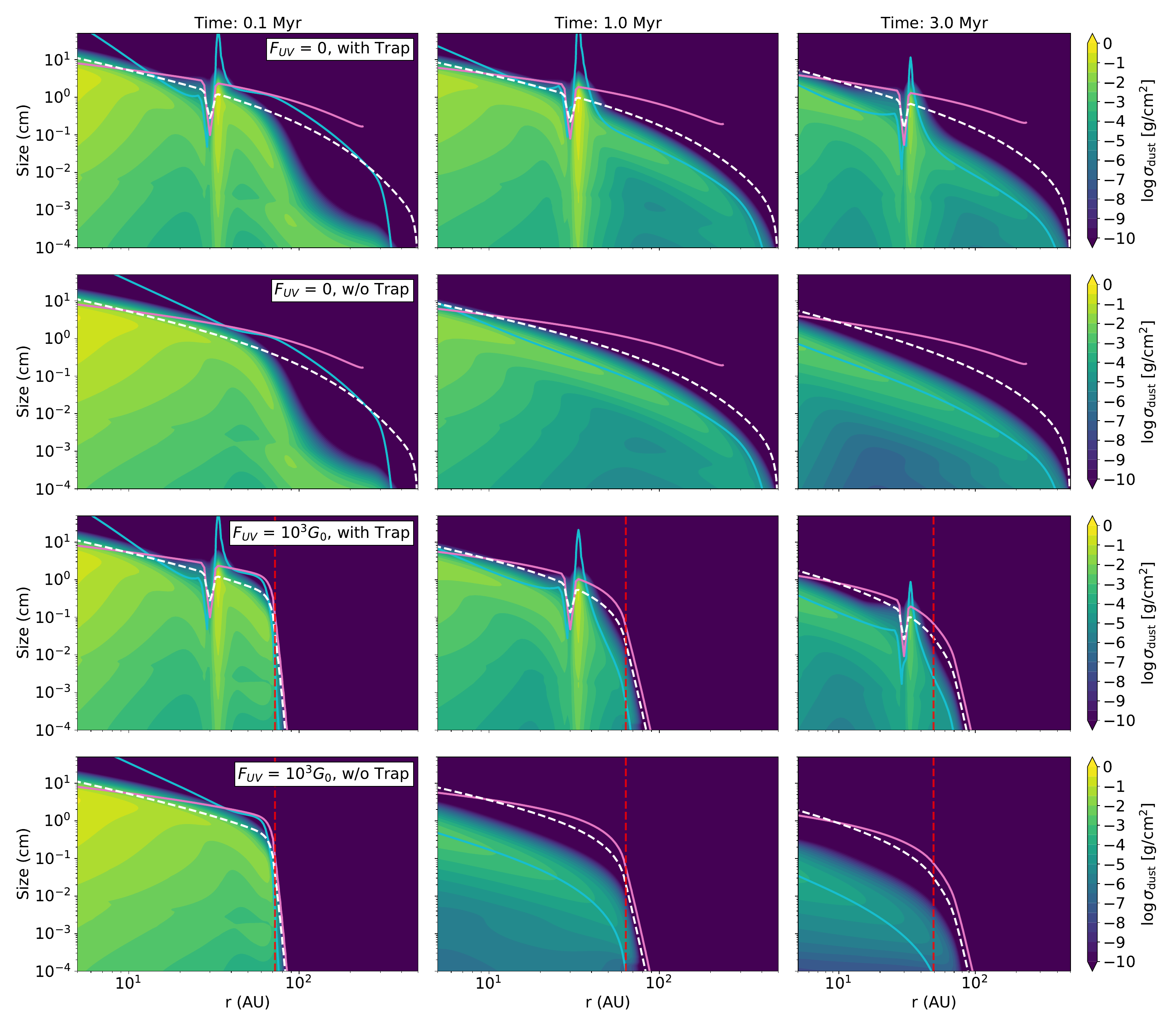}
 \caption{
 Dust size distribution at \SI{0.1}{}, \SI{1}{}, and \SI{3}{Myr} (from left to right), for the simulations with/without photoevaporation, and with/without dust traps.
 The solid lines indicate the estimated fragmentation and drift growth limits (magenta and cyan, respectively). For reference, we show the grain sizes that correspond to $\mathrm{St} = 0.1$ with a dashed white line, and the photoevaporation radius with a dashed red line.
 Note that we plot $\sigma_\textrm{dust}$, which has surface density units, but it is normalized by the logarithmic bin size \citep[see][]{Birnstiel2010}.
 }
 \label{Fig_Demo_DustDist}
\end{figure*}

\begin{figure}
\centering
\includegraphics[width=90mm]{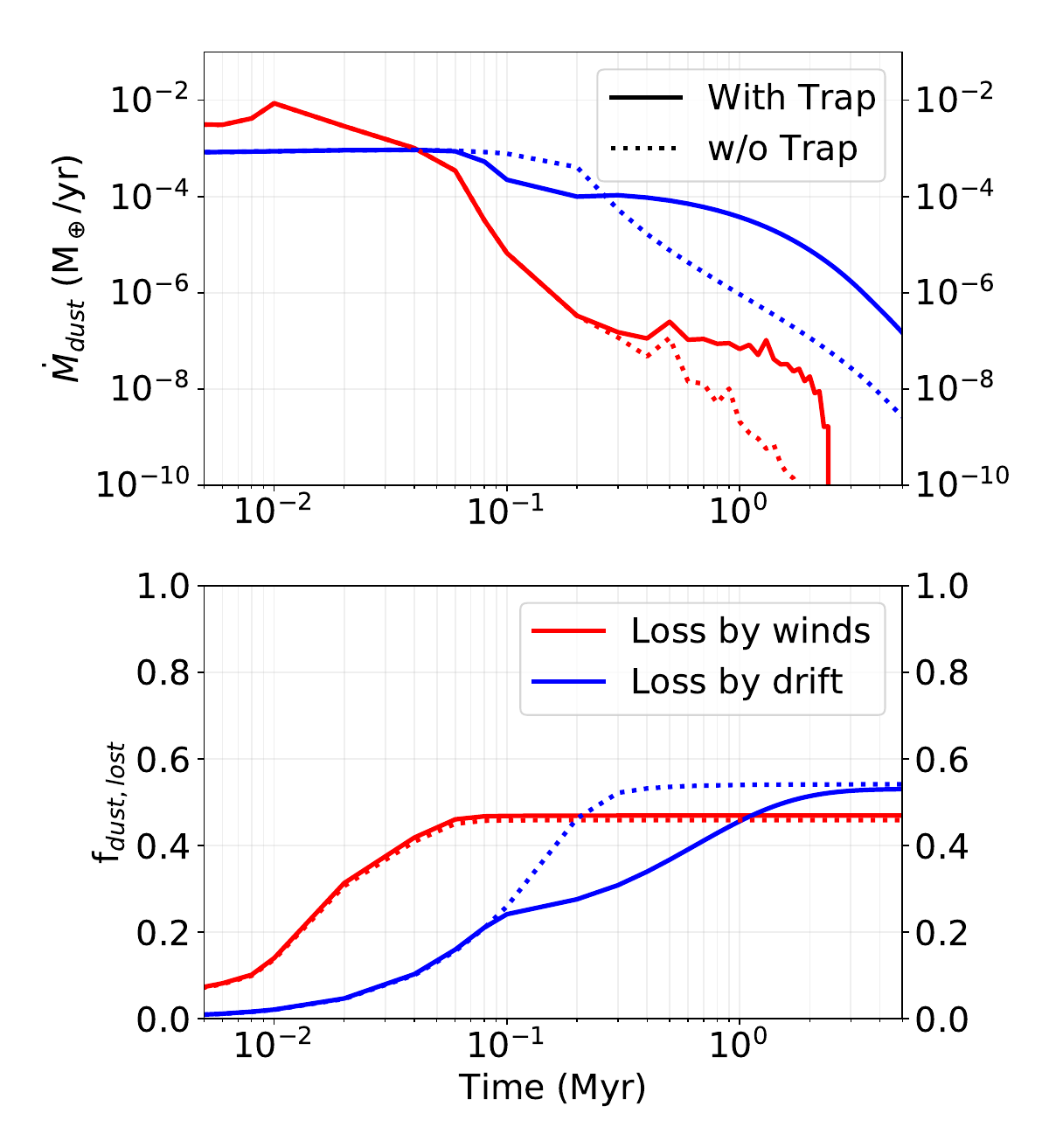}
 \caption{
 \textit{Top:} Evolution of the dust loss rate over time for a photoevaporating disk (with $F_\textrm{UV} = 10^3\, G_0$), with and without dust traps (solid vs. dotted lines). The figure shows the distinction between the dust loss rate due to drift into the star (red), and due to entrainment with the photoevaporative wind (blue).
 \textit{Bottom:} Cumulative fraction of dust lost (relative to the initial dust mass) due to winds and drift.
 }
 \label{Fig_Demo_LossComparison}
\end{figure}

\begin{figure}
\centering
\includegraphics[width=90mm]{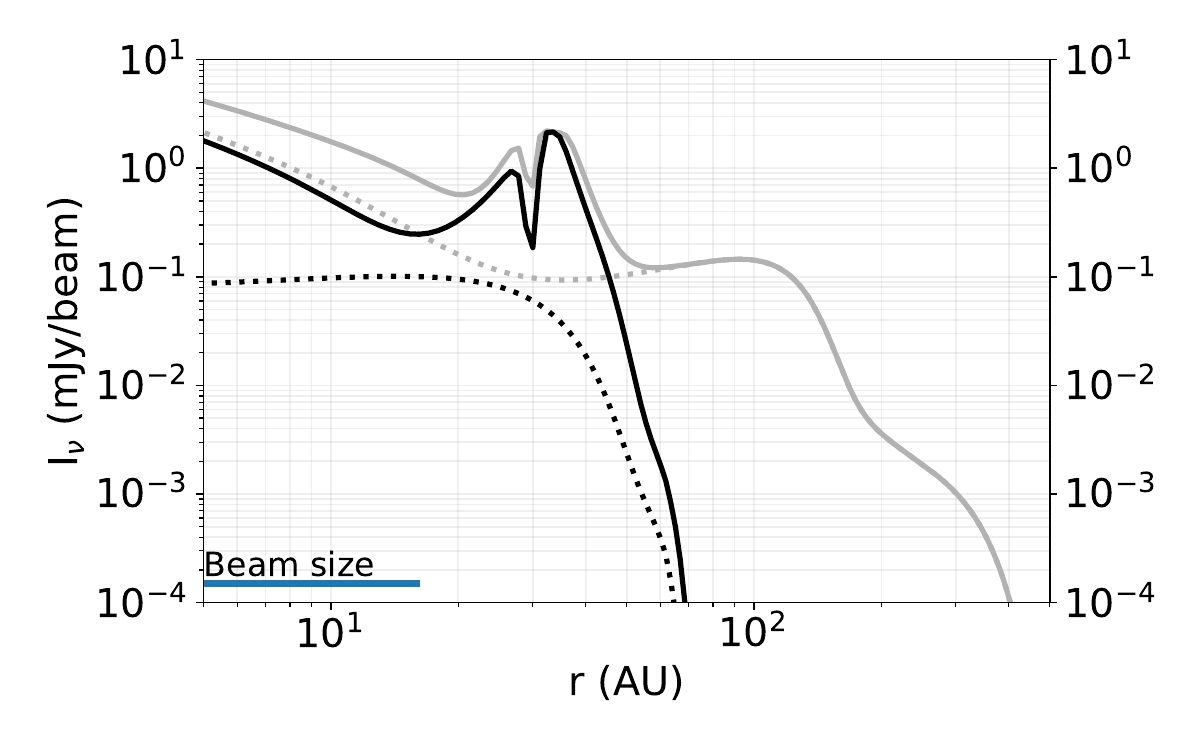}
\includegraphics[width=90mm]{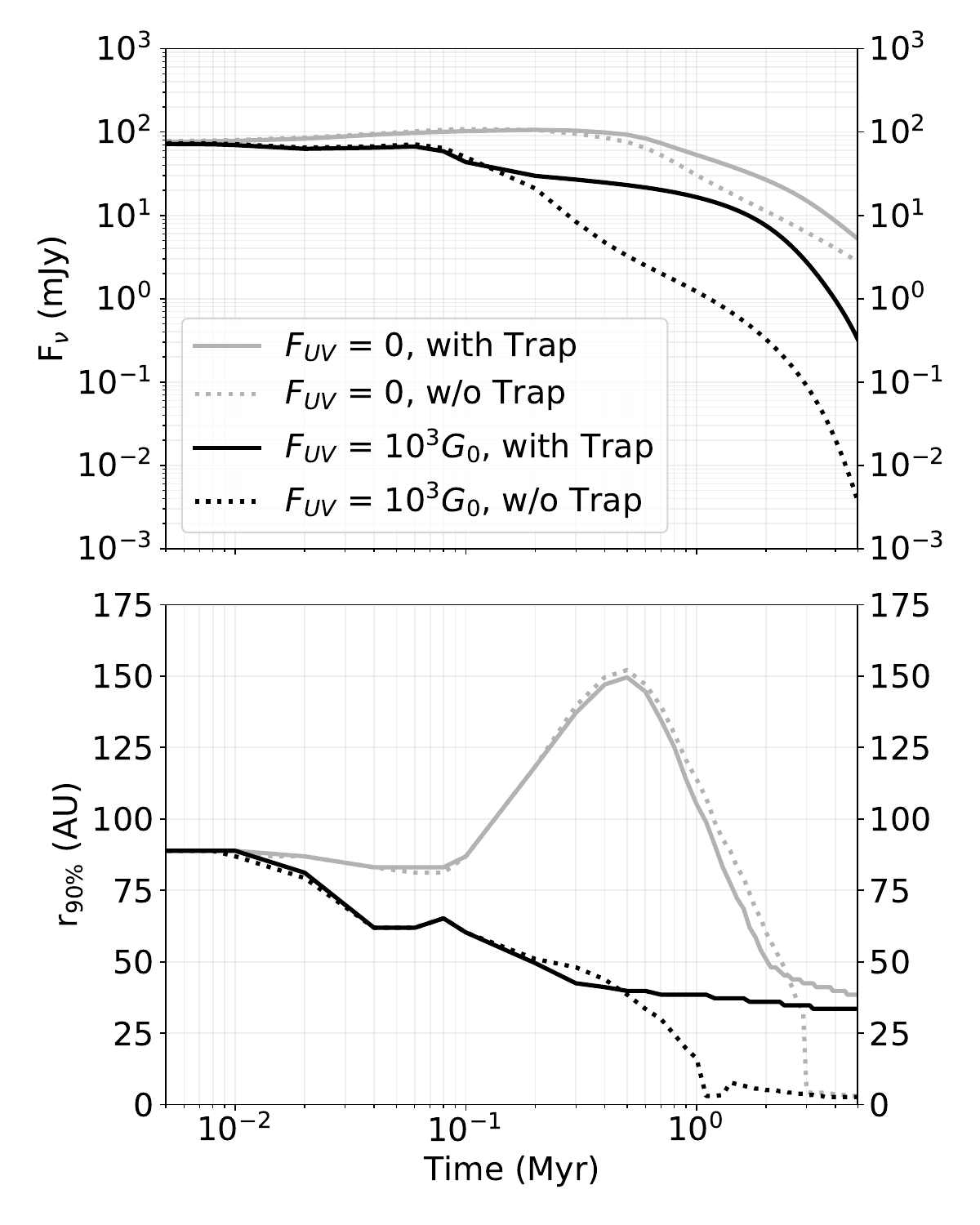}
 \caption{
 \textit{Top:} Intensity profile for the fiducial simulations at $\lambda = \SI{1.3}{mm}$, assuming a distance of $d = \SI{400}{pc}$, and a beam size of $\SI{40}{mas}$ ($\SI{16}{AU}$, shown as a horizontal blue line), for a time of $t = \SI{1}{Myr}$.
 \textit{Middle:} Evolution of the disk flux at $\lambda = \SI{1.3}{mm}$.
 \textit{Bottom:} Evolution of the radius enclosing $90\%$ of the disk continuum emission, assuming a sensitivity threshold of $\SI{0.1}{mJy\, beam^{-1}}$.
 }
 \label{Fig_Demo_IntensityProfile}
\end{figure}

\begin{table*}[h]
\centering
\caption{Dust mass evolution - Fiducial Comparison.}
\label{Table_DustMass}
\begin{tabular}{l l |c c c|c c c|c c c}
 \hline \hline
 & &  & $M_\textrm{dust}\, (M_\oplus)$ & & & $F_\nu$ (mJy) -- $\SI{1.3}{mm}$ & & & $r_{90\%}$ (AU) & \\
$F_\textrm{UV}\, (G_0)$ & Dust trap & \SI{1}{Myr} &\SI{3}{Myr} & \SI{5}{Myr} & \SI{1}{Myr} &\SI{3}{Myr} & \SI{5}{Myr} & \SI{1}{Myr} &\SI{3}{Myr} & \SI{5}{Myr}\\
 \hline
 $0$ &  Yes    & 104.3 & 13.8 & 4.7 & 53.2 & 15.0 & 5.2 & 105.2 & 42.5 & 38.5\\
 $0$ &  No     & 26.8  &  5.6 & 2.9 & 30.5 & 6.3  & 2.8 & 113.9  & < 5 &  < 5 \\
 $10^3$ &  Yes & 22.8  &  1.3 & 0.1 & 16.5 & 2.8  & 0.3 & 38.5 & 34.7 & 33.5\\
 $10^3$ &  No  &  0.4  & 0.02 & 0.003  &  1.2 & 0.1  & 0.004 & 16.2 & < 5 & < 5\\
\hline
\end{tabular}
\vspace{-3.0mm}
\end{table*}

We begin our comparison between simulations with and without external photoevaporation, in order to understand the main features of each evolution pathway. The effect of photoevaporation on the gas component is straightforward (\autoref{Fig_Demo_MassEvolution}, top panel). 
The gas mass decreases faster for disks undergoing  external photoevaporation than in the viscous evolution counterparts, and (for our model) the presence of substructures have no significant impact on the overall gas evolution. Meanwhile, the dust component is distinctively affected by both the effects of photoevaporation and dust traps. \autoref{Fig_Demo_DustDist} shows the morphology of the dust size distribution at relevant times, where we see how photoevaporation truncates the outer disk, and how dust traps retain higher concentrations of large grains.

Disks subject to external photoevaporation have lower dust masses than both of their viscous evolution counterparts (\autoref{Fig_Demo_MassEvolution}, middle panel), and the mass loss in (externally) photoevaporating disks can be described in two stages: a wind dominated loss, and a drift dominated loss (\autoref{Fig_Demo_LossComparison}). During the first stage of disk evolution, the dust grains in the outer regions are still small and can be easily entrained with the photoevaporative winds. 
This leads to a decrease in the dust mass from the initial $\SI{300}{M_\oplus}$ to $\SI{80}{M_\oplus}$ during the first \SI{0.1}{Myrs}. In contrast, the dust mass in viscous evolution simulations only decreases to $\SI{250}{M_\oplus}$ during the same period of time.\par

From dust evolution theory, the lifetime of the disk (in terms of the dust component) is on the same order of magnitude as the drift and growth timescales of dust particles at the disk outer edge \citep[][]{Birnstiel2012_b, Powell2017}.
Because photoevaporation removes all the material from the disk outer regions, effectively truncating the disk size, the remaining dust component will drift faster towards the star, in comparison with the viscous counterparts.\par

\autoref{Fig_Demo_LossComparison} shows that in the simulation with photoevaporation and without dust traps, the dust loss rate is first dominated by wind entrainment ($t \lesssim \SI{0.1}{Myr}$), and quickly  becomes dominated by drift ($t \gtrsim \SI{0.1}{Myr}$). For this simulation, the disk has lost $99\%$ of the initial dust mass by $t_\textrm{depletion} \approx \SI{0.4}{Myrs}$. Up until this point, our results agree with the work of \citet[][see their Fig. 5]{Sellek2020}, and we refer to their paper for a detailed analysis of the evolution of disks without substructures.\par 

It is during the drift dominated stage that the presence of a dust trap becomes relevant, greatly delaying the depletion of the dust component. 
In this simulation with external photoevaporation and dust traps, the total mass lost by drift becomes comparable to that lost by winds after $t = \SI{1}{Myr}$, and the disk loses the $99\%$ of the initial dust mass only by $t_\textrm{depletion} = \SI{2.3}{Myrs}$. This depletion timescale exceeds the values reported in \citet[][]{Sellek2020} by an order of magnitude, indicating that the presence of dust traps may explain why disks in dense star clusters subject to high $F_\textrm{UV}$ fluxes are still detectable in observations \citep{Guarcello2016, Eisner2018, Otter2021}.\par

In terms of the global dust-to-gas mass ratio (\autoref{Fig_Demo_MassEvolution}, bottom panel) it is interesting to note that this is always below the initial $\epsilon_0 = 0.01$, despite the gas mass loss in disks with external photo-evaporation. In particular, the simulation without dust traps under photoevaporation displays a remarkably lower dust-to-gas ratio than its viscous counterpart, which is due to the reduced drift timescale by the photoevaporative truncation of the disk.\par
The effect of photoevaporation and the dust traps can also be seen in the derived observable quantities, i.e. the intensity profiles, the flux, and the dust disk radius that is usually assumed as the radius enclosing $90\%$ of the total millimeter flux (obtained following \autoref{eq_Flux_OpticallyThin}), which are shown in \autoref{Fig_Demo_IntensityProfile} for the wavelength $\lambda = \SI{1.3}{mm}$. We also indicate the dust masses, the corresponding fluxes, and the radii containing 90\% of the dust emission at specific times in \autoref{Table_DustMass}. \par
From the intensity profiles (top panel, shown at \SI{1}{Myr}) we see how the disk subject to external photoevaporation are truncated at $r \approx \SI{60}{AU}\, -\, \SI{70}{AU}$, in contrast with their viscous evolution counterparts, which display a more extended emission.
As expected from the difference in dust masses, the disks with the  gap-substructures are brighter than their smooth counterparts, featuring a ring-like emission at $r \approx \SI{35}{AU}$. 
We note that, since the dust trap is located well inside the truncation radius, the morphology of the bright ring  is not affected by external photoevaporation. \par
The evolution of the total flux $F_\nu$ (middle panel) follows the trend of the total dust mass. Here, the simulations with photoevaporative loss show fluxes that range between \SI{0.01}{}--\SI{15}{mJy} after $t = \SI{1}{Myrs}$, and in particular the simulation with a dust trap manages to display an emission higher than $\SI{1}{mJy}$ until \SI{4}{Myrs}. The value of $r_{90\%}$ (bottom panel) reflects the extension of the intensity profile until the sharp drop below the sensitivity limit ($\SI{0.1}{mJy\, beam^{-1}}$ for this plot). 
For the disks subject to external photoevaporation the disk size progressively shrinks, first until the truncation radius within the first $\SI{0.1}{Myr}$ due to the initial dust removal by winds. Then, if no dust traps are present, the disk continues to shrink until it disappears at $t\approx \SI{1}{Myr}$ (i.e., the emission is fully below the sensitivity limit) or, if substructures are present, the disk only shrinks until the location of the dust ring.\par

We also note that the viscous evolution simulations first experience an expansion until reaching $r_{90\%} = \SI{150}{AU}$ at $t = \SI{0.5}{Myr}$, and then contract as the dust drifts inwards, as the emission from outer regions of the disk falls below the sensitivity limit. At $\SI{3}{Myr}$ the dust disk sizes in the viscous simulations either reach the dust trap location or drop sharply (for the case without substructures).\par

\subsection{Parameter Space: UV field and Trap location}\label{sec_Results_ParamSpace}

\begin{figure*}
\centering
\includegraphics[width=180mm]{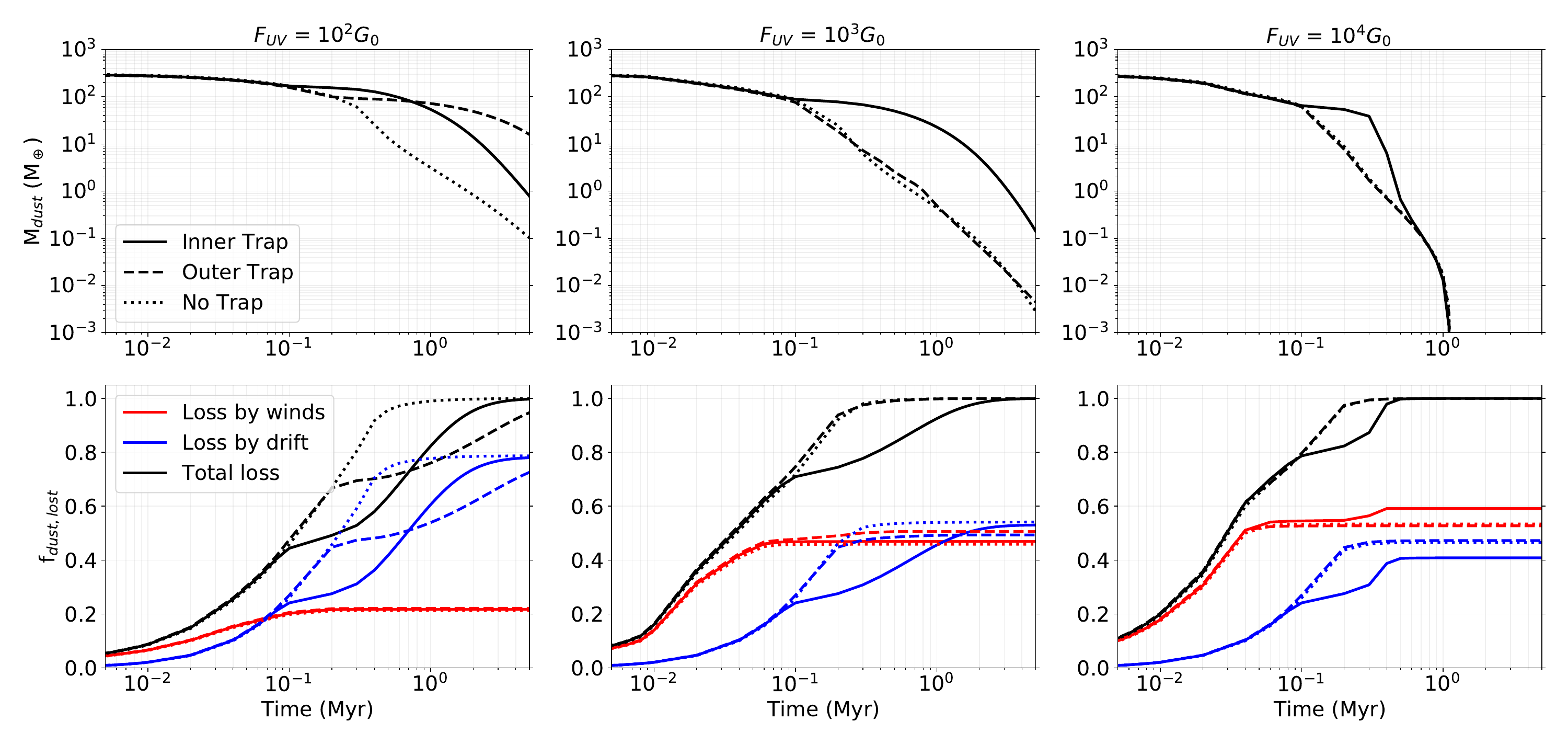}
 \caption{
 \textit{Top:} Mass evolution for our parameter space in UV fluxes and trap location (or absence).
 \textit{Bottom:} Fraction of dust mass lost. The lines distinguish between the total dust lost (black), the fraction lost by drift at into the star (blue), and lost by entrainment with the photoevaporative wind (red). 
 }
 \label{Fig_Key_MassEvolution}
\end{figure*}

\begin{figure*}
\centering
\includegraphics[width=180mm]{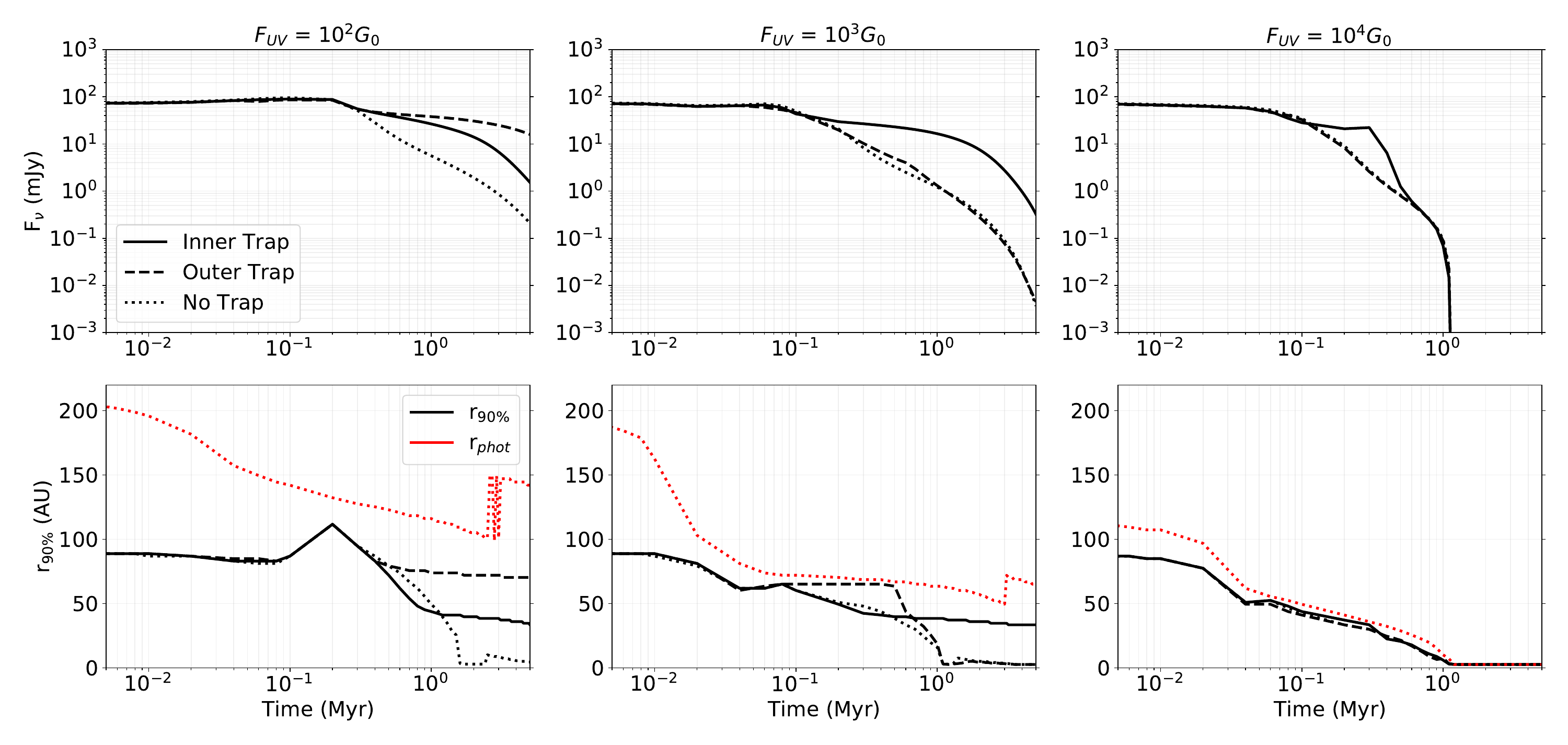}
 \caption{
 Evolution of the disk observables, for different FUV fields and trap locations ($1/3 r_c$, $2/3 r_c$ or none).
 \textit{Top:} Evolution disk flux in the \SI{1.3}{mm} continuum at a distance of \SI{400}{pc}.
 \textit{Bottom:} Time evolution of the $r_{90\%}$ enclosing $90\%$ (black) of the disk emission in the continuum above the sensitivity limit ($0.1$ mJy beam$^{-1}$), and the photoevaporative radius derived from the FRIED grid mass loss profile \citep[red dotted lines, see][]{Sellek2020}. We note that towards the end of the simulation the calculation of the photoevaporative radius suffers from numerical effects due to the discrete nature of the FRIED grid.}
 \label{Fig_Key_ObservableEvolution}
\end{figure*}

\begin{table}[h]
\caption{Dust depletion timescales. "Dispersed" refers to dust traps that have vanished in the simulation due to external photoevaporation.}
\label{Table_DispersedTimescale}
\begin{tabular}{c c c }
 \hline \hline
$F_\textrm{UV}\, (G_0)$ & Trap Location &$t_\textrm{depletion}$ (Myr)\\ 
 \hline 
        &  Inner    &  3.3  \\
 $10^2$ &  Outer    &  >5.0 \\
        &  None     &  1.0  \\ \hline
        &  Inner    &  2.3  \\ 
 $10^3$ &  Outer (dispersed)    &  0.4  \\
        &  None     &  0.3  \\  \hline
        &  Inner (dispersed)    &  0.4  \\
 $10^4$ &  Outer (dispersed)   &  0.2  \\
        &  None     &  0.2  \\
\hline
\end{tabular}
\vspace{-3.0mm}
\end{table}

In this section, we study how the presence of dust traps affects the dust mass and observable properties of the disk depending on the external FUV flux (\autoref{Fig_Key_MassEvolution} and \autoref{Fig_Key_ObservableEvolution}), and also list the dust depletion timescale (i.e. the age of the disk when $99\%$ of the initial dust mass has been lost) covering the parameter space of \autoref{Table_DispersedTimescale}. The table also indicates whether a dust trap was dispersed by the photoevaporative winds within the $\SI{5}{Myr}$ of simulation time.\par
For the strongest FUV flux ($10^4\, G_0$), the mass loss due to photoevaporation is so intense, that it renders existing substructures irrelevant in terms of observability.
From the evolution of $r_{90\%}$ and the photoevaporative radius we can see how photoevaporation progressively shrinks the disk, until it completely disappears by $\SI{1}{Myr}$, approximately at the same time that the emission in the millimeter continuum drops. 
We see that if the substructure is located in the inner regions ($r_\textrm{gap} = \SI{30}{AU}$, or $1/3\, r_c$ in our simulation), there is only a minor delay of approx \SI{0.2}{Myr} until the flux in the millimeter continuum drops.\par
For the medium FUV flux ($10^3\, G_0$), we already saw in the previous section that an inner dust trap can extend the disk lifetime by $\SI{2}{Myr}$, and that the disk $r_\textrm{90\%}$ size converges to the location of the dust trap, as it also occurs in models with dust traps but without external photoevaporation \citep[e.g.][]{Stadler2022}.
However, we also find that in the case where the substructure is located further out ($r_\textrm{gap} = \SI{60}{AU}$, or $2/3\, r_c$), the disk would still be dispersed by photoevaporation, once the photoevaporative radius reaches the location of the dust trap. This can be seen in \autoref{Fig_Key_ObservableEvolution} (mid-bottom panel, dashed line), where the disk size initially matches the location of the outer dust trap, and quickly disperses once the disk $r_\textrm{90\%}$ becomes comparable with the photoevaporative radius. In terms of the dispersal timescale and continuum fluxes, there is almost no difference between having an outer dust trap or no dust trap at all.\par
In the case with a weak FUV field ($10^2\, G_0$), both the inner and outer dust trap remain inside the photoevaporation radius 
(as seen from the value of $r_{90\%}$), which means that the size and flux of the disk will be dominated by the material trapped at local the pressure maximum. 
In comparison with the medium and strong photoevaporative regimes, disks in the weak photoevaporative environment have longer lifetimes of $t_\textrm{depletion} = \SI{3}{Myr}$ and $t_\textrm{depletion} \gtrsim \SI{5}{Myr}$, for the cases with inner and outer substructures  respectively. 
For this regime, the lifetime of the disk with the outer substructure is longer than the one with the inner substructure, 
which we infer to be due to the longer drift timescales from the outer dust trap to the star, and the longer diffusion timescales across the gap in the outer regions.\par

From these three photoevaporative regimes we conclude that the lifetime of the disk is dominated by the outermost dust trap, provided that this is located well inside the photoevaporative radius, and that the dust trap located in the photoevaporative regions (even marginally) would have little to no effect on the disk survival timescale. It might seem surprising that disks with fully formed dust traps are dispersed, since photoevaporative winds can only carry away small grains. However, because fragmentation of large grains continuouslyreplenishes the population of micron-sized particles, there is always a non-negligible fraction of solids being carried away. We will revisit the dust size distribution of entrained grain in \autoref{sec_Results_LostGrains}.\par

\subsection{Low mass stars and compact disks}

\begin{figure}
\centering
\includegraphics[width=90mm]{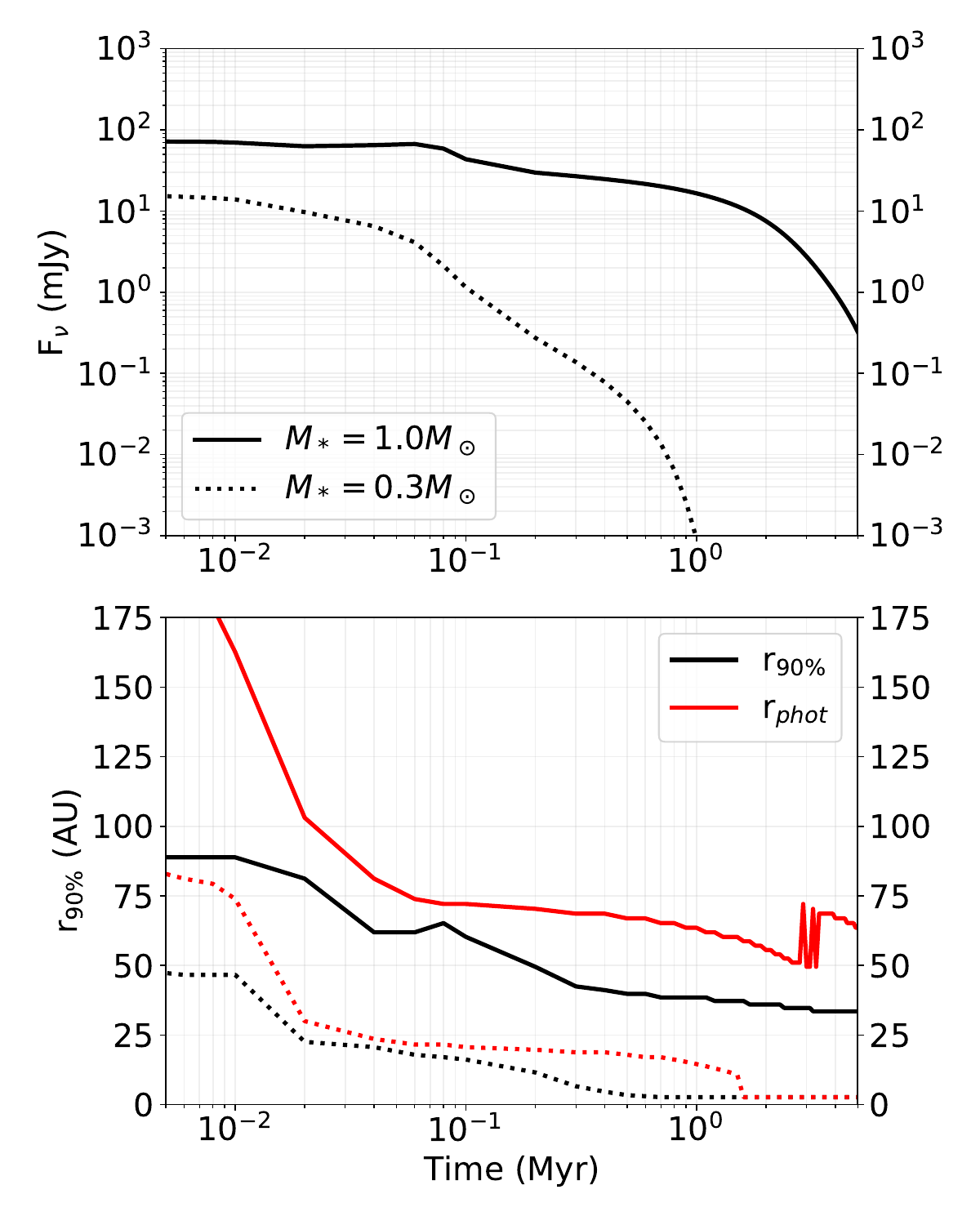}
 \caption{
Evolution of the disk continuum flux at  $\lambda = \SI{1.3}{mm}$ and the $r_{90\%}$ (assuming a sensitivity threshold of $\SI{0.1}{mJy\, beam^{-1}}$) for two different stellar masses, and including the gap at $r_c = \SI{30}{AU}$. The photoevaporative radius is shown for comparison. The UV flux is $10^3\, G_0$.}
\label{Fig_StellarMassComparison}
\end{figure}

\begin{figure}
\centering
\includegraphics[width=90mm]{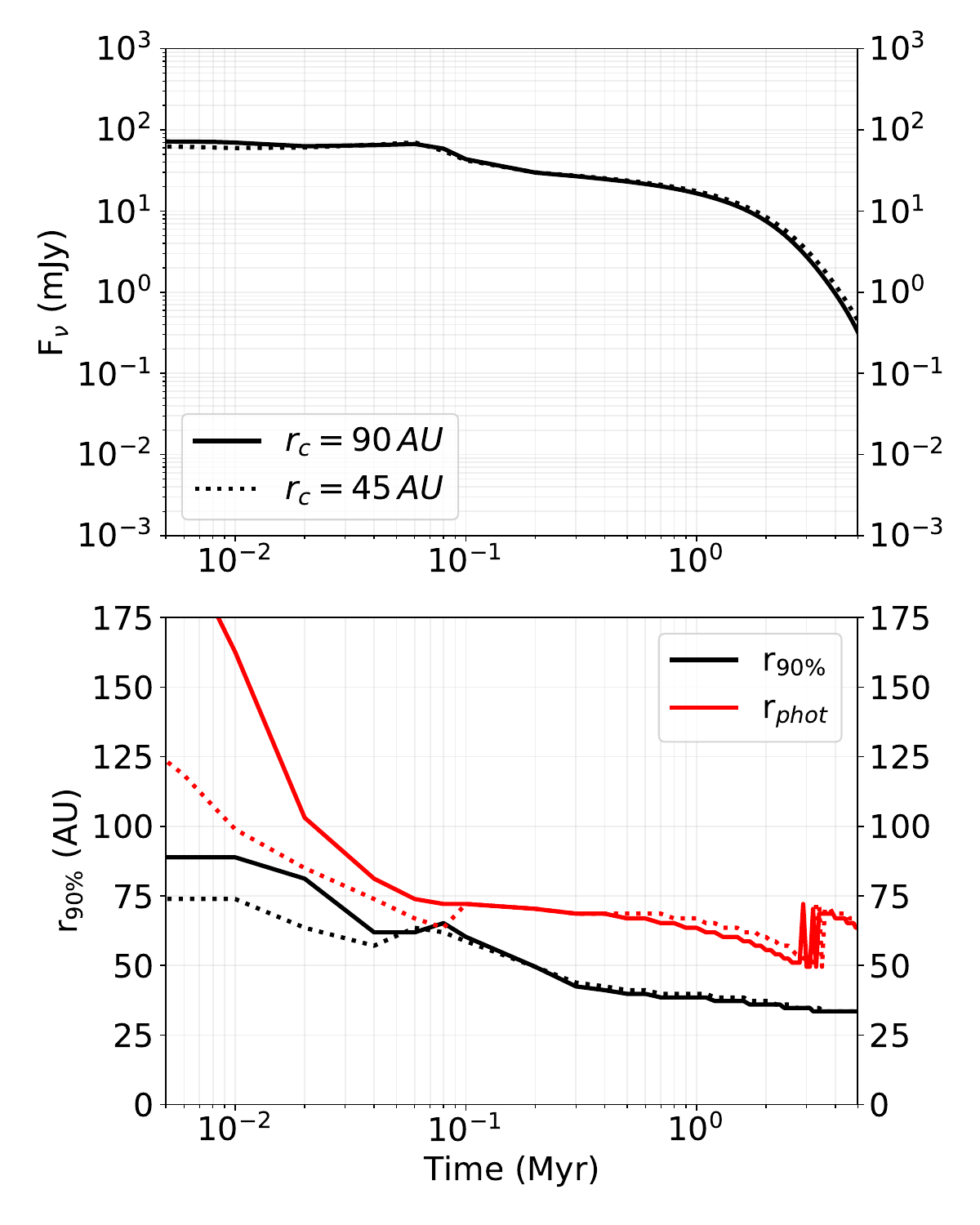}
 \caption{
Same as \autoref{Fig_StellarMassComparison} for two initial characteristic radius $r_c$.  The UV flux is $10^3\, G_0$.
}
\label{Fig_DiskRcComparison}
\end{figure}

We perform two additional simulations to study whether a dust trap would survive in a disk around a low mass star ($M_* = \SI{0.3}{M_\odot}$), and if an initially more compact disk size ($r_c = \SI{60}{AU}$) would affect our observational predictions. For this comparison, we use a radiation field of $F_{UV} = 10^3\, G_0$, and fix the dust trap location at $\SI{30}{AU}$.\par
In \autoref{Fig_StellarMassComparison} we see that the disk around a $\SI{0.3}{M_\odot}$ mass star is dispersed faster than the one around a sun-like star, and that even the presence of the dust trap cannot prevent the sharp decrease in the millimeter continuum flux.
This faster dispersal occurs because the gravitational potential is weaker around a low mass star, which means that the gas and solid can be removed more easily by the UV irradiation from the environment. Consequently, the photoevaporation radius can reach further into the inner regions of the disk, all the way down to $r_\textrm{phot} \approx\SI{20}{AU}$, and disperse the dust trap that was located at $\SI{35}{AU}$. In order for the dust component to survive in a disk around low mass star, the trap should be located further inward ($r_\textrm{gap} \lesssim \SI{10}{AU}$), provided that the UV field is on the order of $F_{UV} = 10^3 G_0$. Disks in regions with lower irradiation could still display dust traps at larger radii.\par

For the case of a disk that is initially more compact, we do not see any significant differences in the disk evolution in terms of the observed size $r_{90\%}$ or the flux in the millimeter continuum (see \autoref{Fig_DiskRcComparison}). From this plot, we can expect for disk sizes in photoevaporative regions to be determined by their outermost dust trap, independently of the initial extension.\par

\subsection{Variable UV field} \label{sec_Results_VariableUV}

\begin{figure}
\centering
\includegraphics[width=90mm]{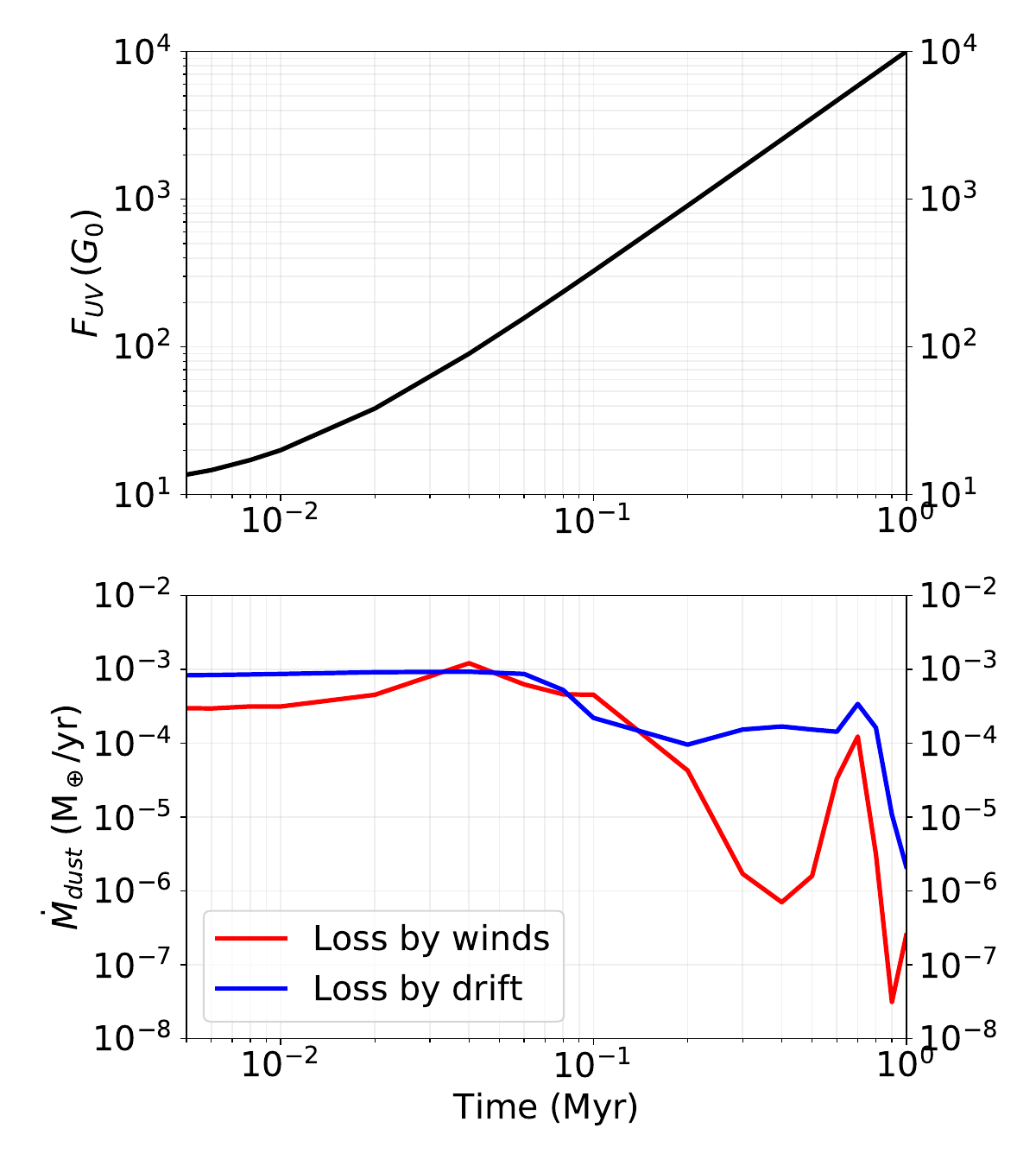}
 \caption{
\textit{Top:} Time evolution of the UV flux following \autoref{eq_IncreasingFlux}.
\textit{Bottom:} Evolution of the dust mass loss rate by drift (blue) and photoevaporative winds (red) in a disk with variable UV flux. The stellar mass and initial size correspond to the fiducial model ($M_* = \SI{1}{M_\odot}$, $r_c = \SI{90}{AU}$).
}
\label{Fig_UVInc_MassLoss}
\end{figure}

\begin{figure}
\centering
\includegraphics[width=90mm]{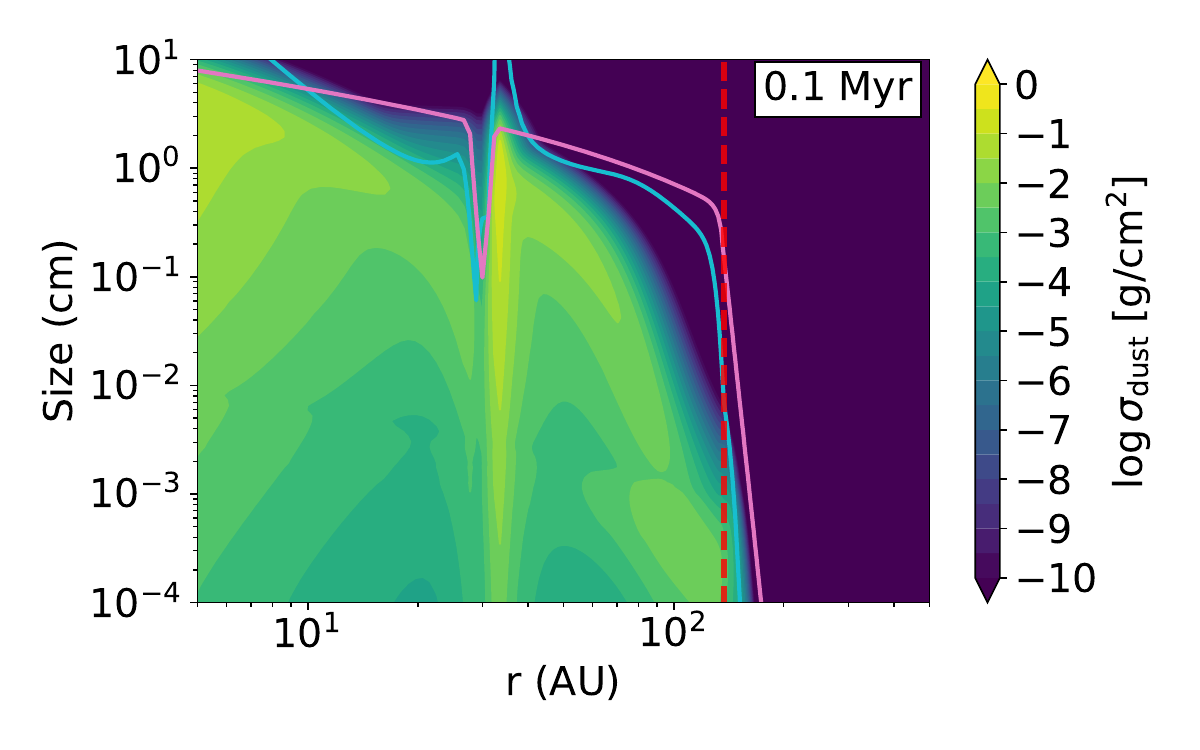}
\includegraphics[width=90mm]{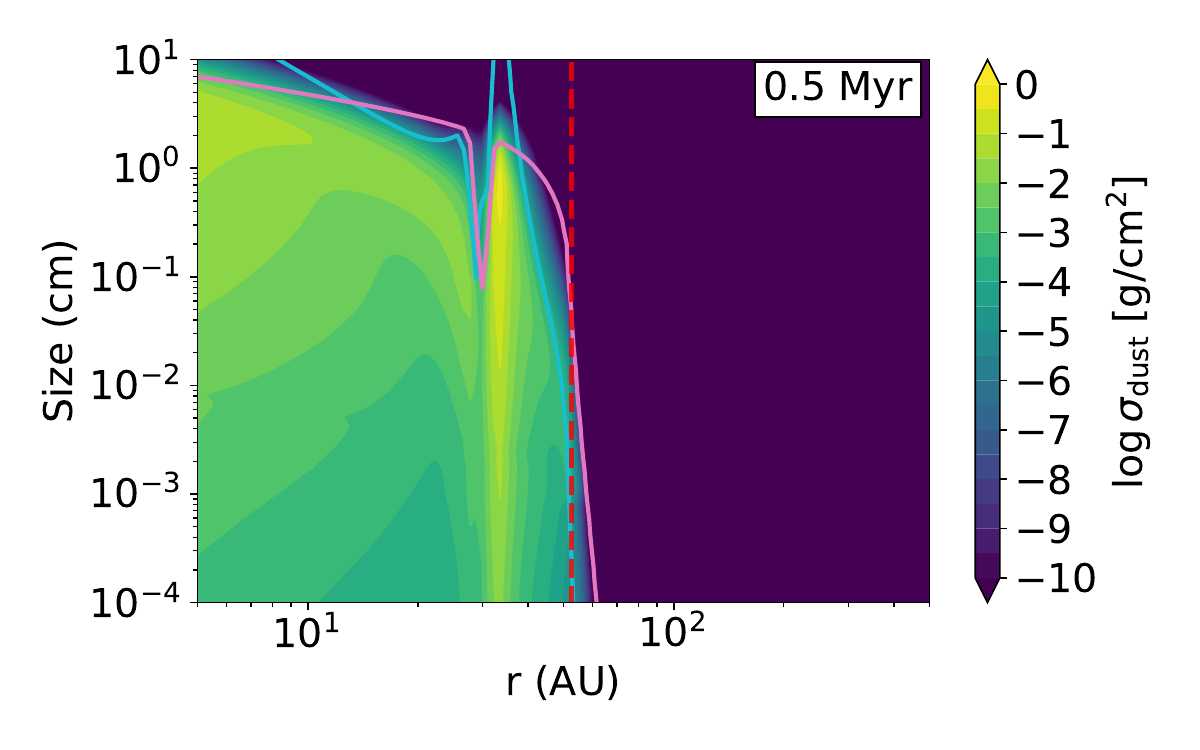}
\includegraphics[width=90mm]{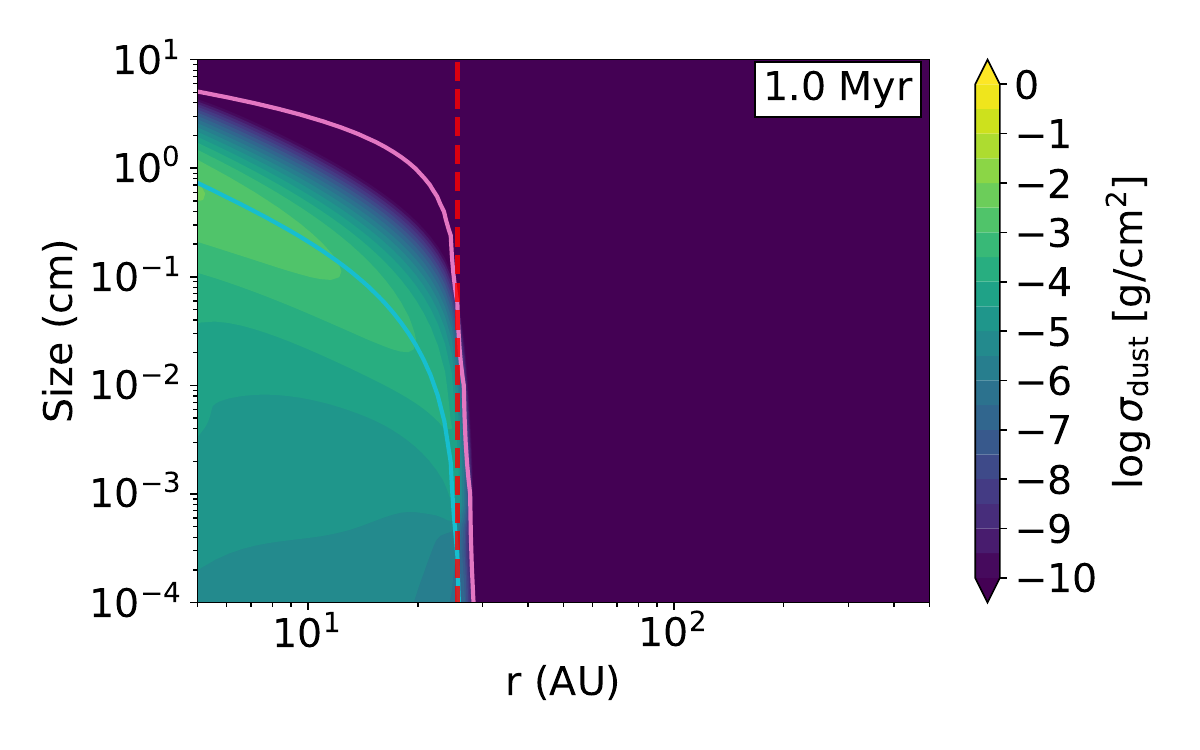}
 \caption{
 Dust size distribution at \SI{0.1}{}, \SI{0.5}{}, and \SI{1}{Myr} for the simulation with an increasing UV Flux (\autoref{eq_IncreasingFlux}). As in \autoref{Fig_Demo_DustDist}, the fragmentation limit is marked in magenta, the drift limit in cyan, and the photoevaporation radius as a vertical dashed-red line. 
 }
 \label{Fig_UVInc_DustDist}
\end{figure}

\begin{figure}
\centering
\includegraphics[width=90mm]{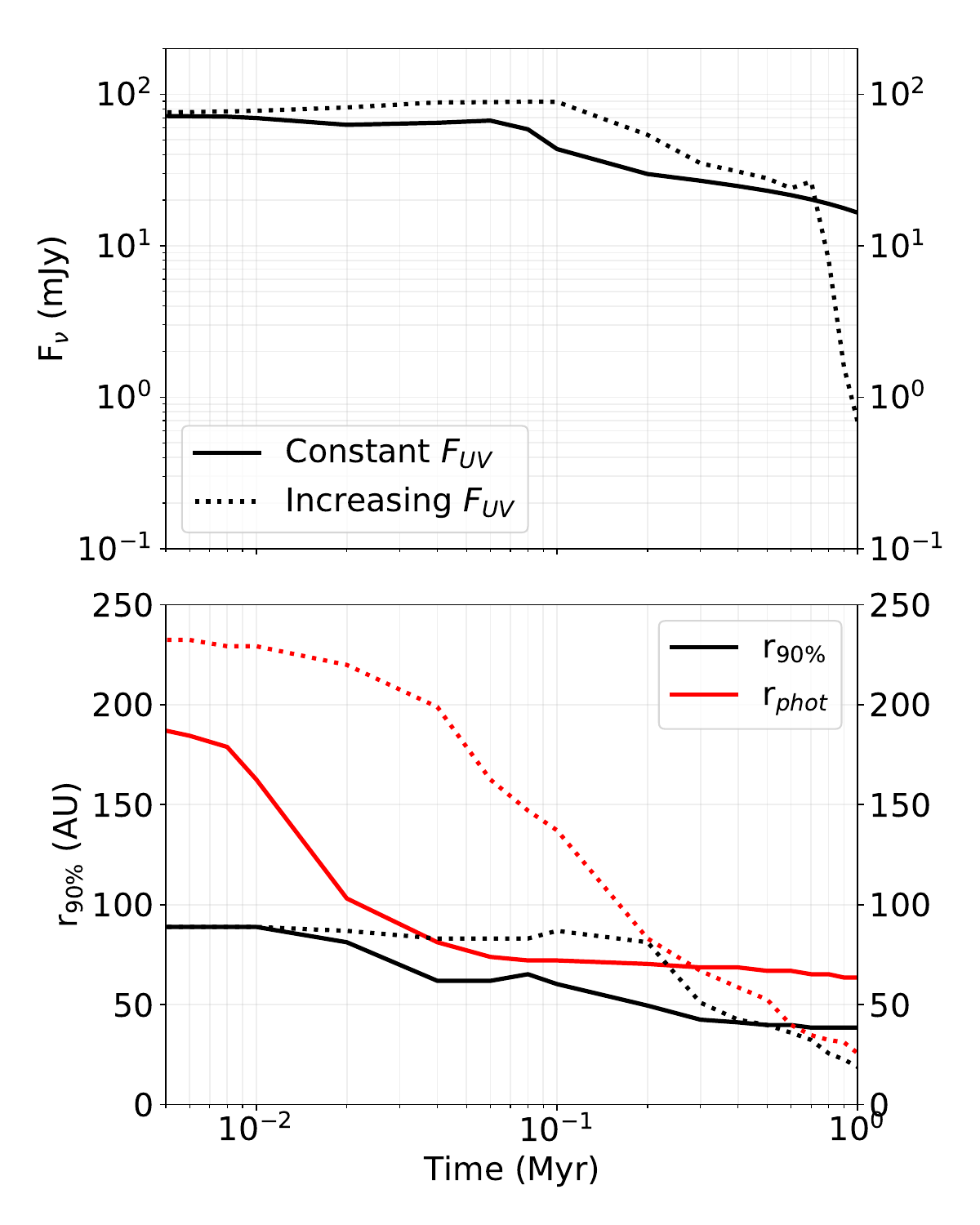}
 \caption{
Same as \autoref{Fig_StellarMassComparison}, comparing the simulation with constant UV Flux of $10^3\, G_0$ and the one with increasing UV flux (\autoref{eq_IncreasingFlux}).
}
\label{Fig_UVInc_Comparison}
\end{figure}

\cite{Winter2019} showed some of the disks in high irradiation regions, could have actually formed and migrated from lower irradiation environments. This would explain why these objects are still observable despite the short dispersal timescales associated with the high irradiation.\par

We conduct a simple simulation with our fiducial parameters ($M_* = \SI{1}{M_*}$, $r_c = \SI{90}{AU}$, $r_\textrm{gap} = \SI{30}{AU}$) in which we gradually increase  the $F_{UV}$ irradiation with the following function:

\begin{equation} \label{eq_IncreasingFlux}
    F_{UV} = F_{UV, 0} + (F_{UV,\textrm{max}} - F_{UV,0})  \left(\frac{t}{\SI{1}{Myr}}\right)^{3/2},
\end{equation}
where $F_{UV,0} = 10\, G_0$ and $F_{UV,\textrm{max}} = 10^4\, G_0$ are the lower and upper limits of the FRIED grid \citep{Haworth2018}.
The exponent of $3/2$ is meant to represent an UV flux that increases rapidly as the disk approaches the high irradiation regions, though the exact shape would depend on variables such as the disk trajectory, the distribution of bright massive stars within the cluster \citep{Winter2019, Qiao2022, Wilhelm2023}, and the variation in their luminosity with time \citep{Kunitomo2021}. 
In \autoref{Fig_UVInc_MassLoss} we show the evolution UV flux and the mass loss rate over time up to $\SI{1}{Myr}$.\par

From the dust distribution shown in \autoref{Fig_UVInc_DustDist}  we see how the radial extension of the dust component gets progressively smaller as time passes, and in particular, we see how the dust trap completely vanishes by \SI{1}{Myr} once the UV flux reaches its peak. We note that the dust trap dispersal occurs despite the fact that dust grains have already grown larger into millimeter-to-centimeter sized pebbles (\autoref{Fig_UVInc_DustDist}, middle panel). While these particles are not easily entrained by the wind, they are still being indirectly depleted, since they continuously fragment into the smaller size grains that are directly removed by photoevaporation.\par

This phenomenon can also be seen in the spike in dust loss rate at \SI{0.7}{Myr} (\autoref{Fig_UVInc_MassLoss}, bottom panel), which coincides whith the photoevaporative radius when it catches up with the location of the dust trap, and in the millimeter continuum (\autoref{Fig_UVInc_Comparison}), when the flux sharply drops. 
From our results in \autoref{Table_DispersedTimescale} for the high UV fluxes ($10^4\, G_0$) we can expect the remaining dust component to disperse in timescales of \SI{0.2}{Myr} or less. \par

This implies that even if dust traps manage to survive during an early low irradiation, for example if the disk was initially shielded from UV irradiation \citep{Qiao2022, Wilhelm2023}, farther away from the dense regions of the cluster \citep{Winter2019}, or surrounded by intermediate mass stars ($M_\lesssim 3 M_\odot$) in early evolutionary stages \citep[when the UV emission is lower][]{Kunitomo2021}, once the FUV flux increases the dust component in the dust traps will be quickly dispersed along with the gas.

\subsection{Dust distribution of the lost grains} \label{sec_Results_LostGrains}

\begin{figure}
\centering
\includegraphics[width=90mm]{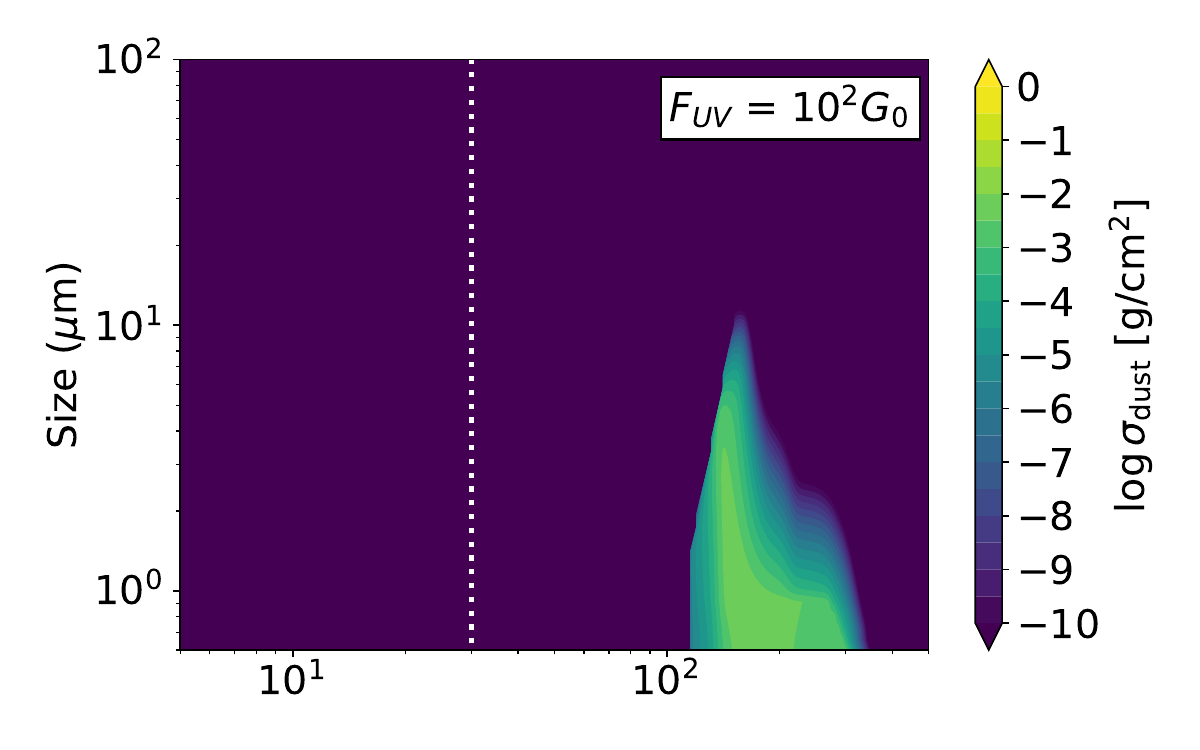}
\includegraphics[width=90mm]{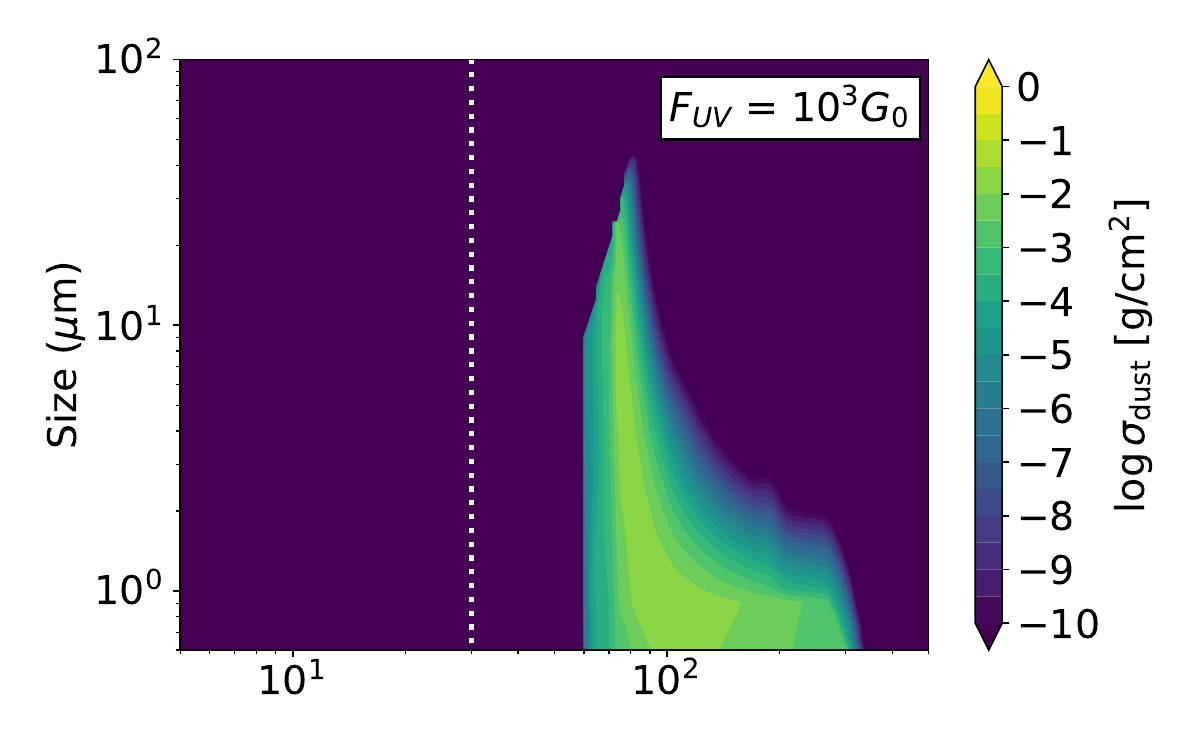}
\includegraphics[width=90mm]{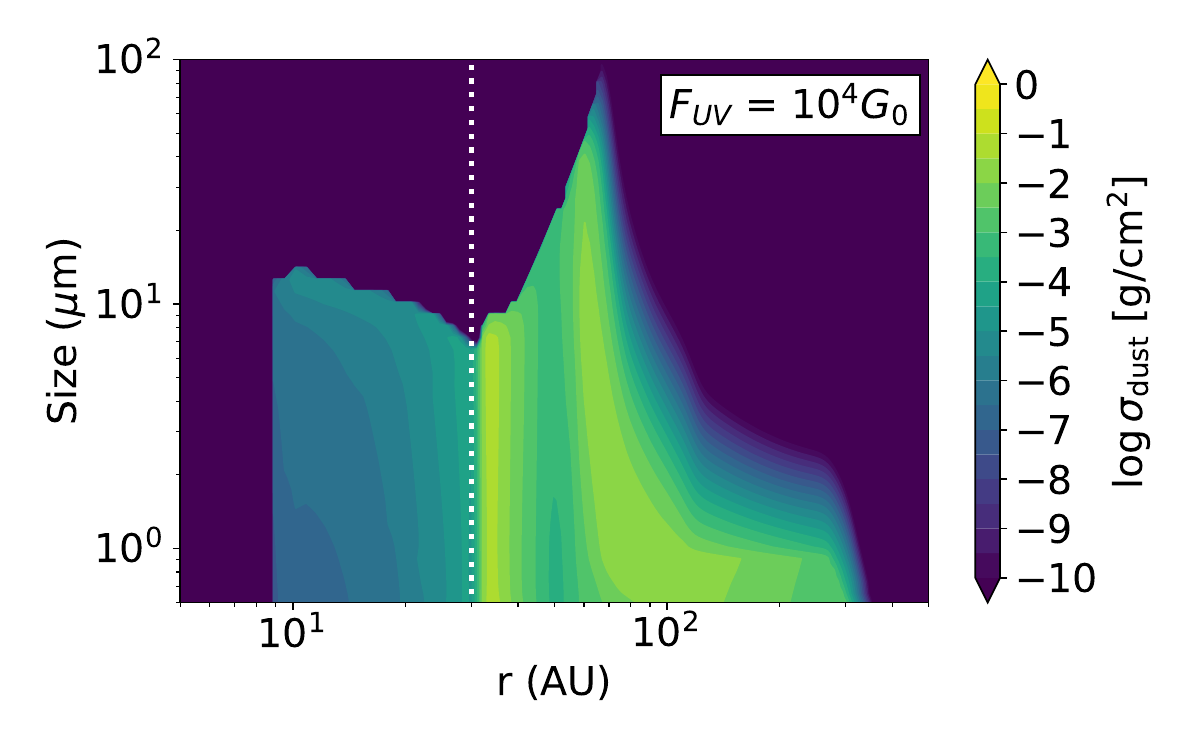}
 \caption{
 Grain size distribution (integrated in time) of the dust grains lost with the photoevaporative flow for the disks subject to low, medium and high $F_{UV}$ environments, at $t=\SI{1}{Myr}$. The disk's parameters are a star of $1 M_\odot$, initial size of $r_c = \SI{90}{AU}$, and a gap substructure at $\SI{30}{AU}$. The location of the gap $r_\textrm{gap}$ is indicated with a dotted white line.}
 \label{Fig_LostDust_SizeDist}
\end{figure}

\begin{figure}
\centering
\includegraphics[width=90mm]{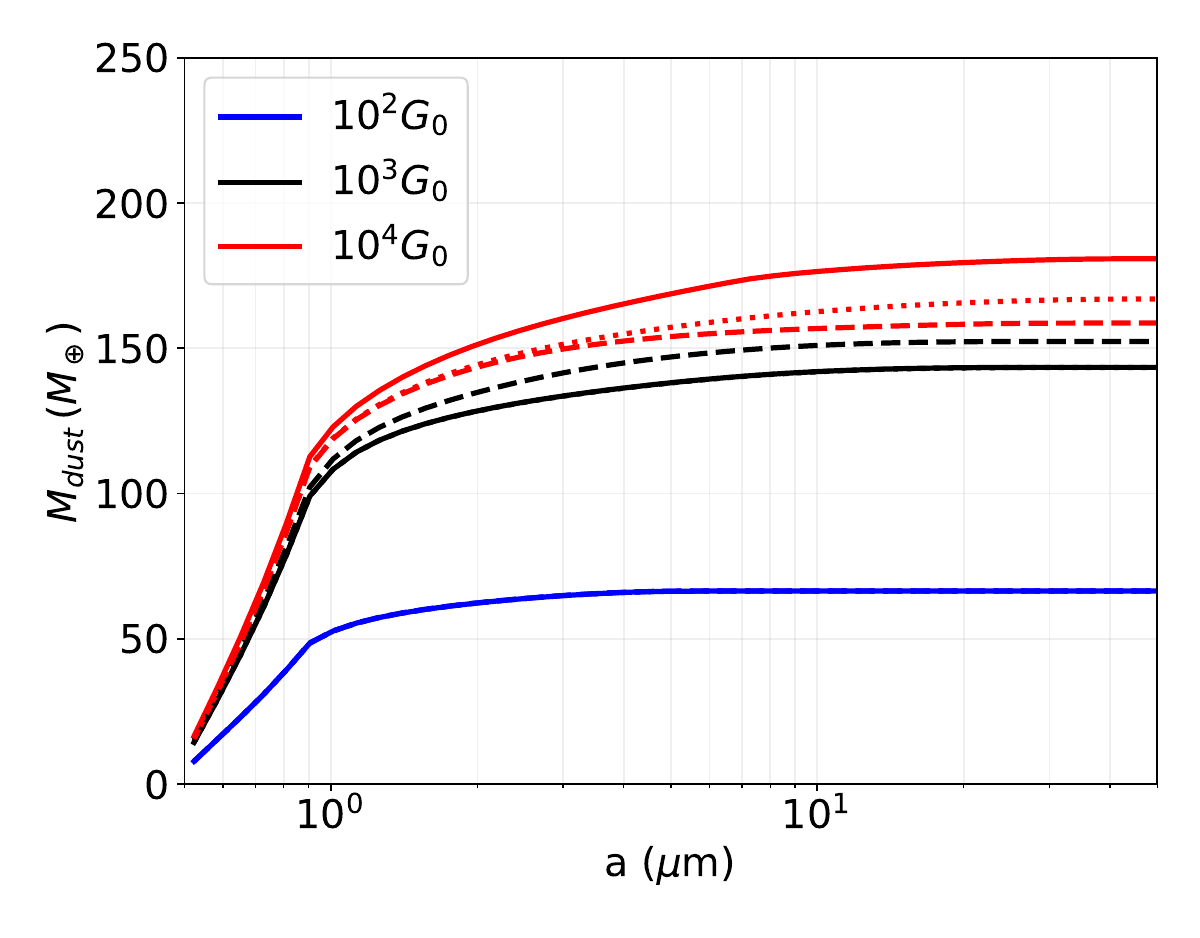}
 \caption{
 Cumulative size distribution of the dust grains lost up to a time of $\SI{1}{Myr}$ for the simulations shown in \autoref{sec_Results_ParamSpace}, for the different UV fluxes, and trap properties. The line styles represent the disks with an inner trap, an outer trap, and no traps at all (solid, dashed, and dotted lines respectively). For the disks with low irradiation environments ($10^2 G_0$) the dust traps have no effect on the lost dust distribution and the three lines overlap.}
\label{Fig_LostDust_CumDist}
\end{figure}

\begin{figure}
	\centering
	\includegraphics[width=90mm]{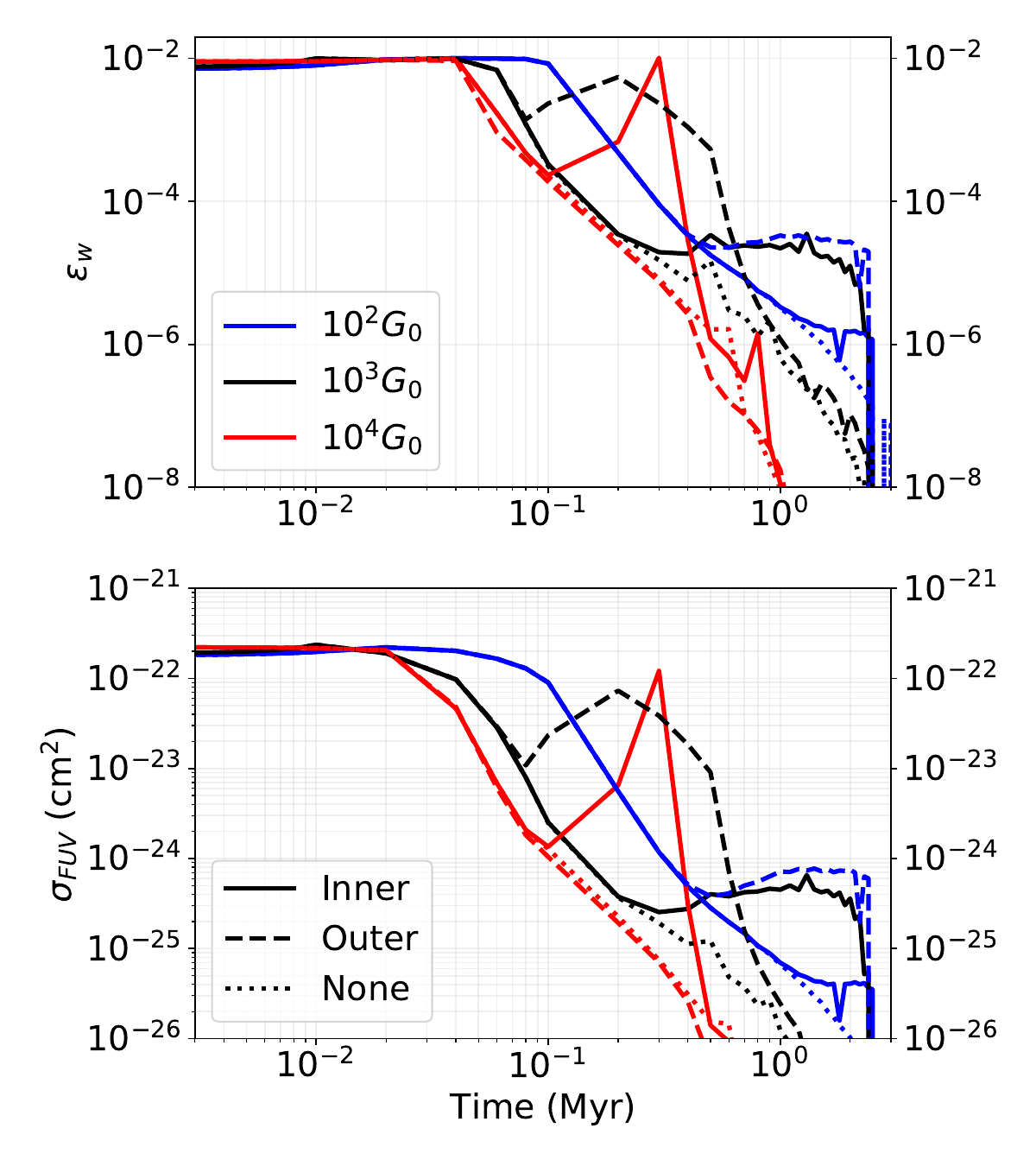}
	\caption{\textit{Top:} dust-to-gas ratio of the material removed by the photoevaporative winds for different irradiation environments and dust trap configurations. \textit{Bottom:} FUV cross section per gas molecule of the material removed by stellar winds, considering the dust-to-gas ratios from the top panel.}
	\label{Fig_LostDust_D2G_Sigma_inst}
\end{figure}

Photoevaporative winds can remove the small grains entrained with the gas flow from the disk outer regions (defined through the photoevaporative radius). 
While tracking the subsequent spatial evolution of these grains goes beyond the scope of this paper, we can still record of the mass and size distribution of lost dust component, and compare between the different irradiation regimes.
\autoref{Fig_LostDust_SizeDist} shows the time integrated distribution of all the dust grains that have been lost by entrainment with the wind up to \SI{1}{Myr}, as a function of the grain size and the launching location, that is:
\begin{equation}
  \Sigma_\textrm{d, lost}(r,a)  = \int_0^{\SI{1}{Myr}} \dot{\Sigma}_\textrm{d,w} (r, a) \stdiff{t},
\end{equation}
where we see how the maximum grain size entrained in the winds and the launching region  depend on the irradiation from the environment.
In particular, we see how for the weakest irradiation field ($10^2\, G_0$) only grains smaller than \SI{10}{\mu m} can be removed from regions beyond \SI{100}{AU}, while for the strongest irradiation field ($10^4\, G_0$) even grains of sizes up to $a \approx \SI{100}{\mu m}$ can be dragged along by the winds, with a noticeable increase in the dust removal at $r \approx \SI{30}{} - \SI{40}{AU}$  where the dust trap is located.\par

To obtain a broader overview of the lost grain distribution, we show in \autoref{Fig_LostDust_CumDist} the cumulative size distribution for the parameter space shown in \autoref{sec_Results_ParamSpace}, i.e.:
\begin{equation}
  M_\textrm{dust, lost}(a)  = \sum^a \int_{r_{in}}^{r_{out}} 2\pi r\,  \Sigma_\textrm{d,lost} (r)\, \stdiff{r},
\end{equation}
which shows that approximately $70\% - 80\%$ of the lost mass of solids comes from the sub-micron size particles. 
Overall, we find that the location of the dust trap has little to no impact in the size distribution of the lost grains (specially in the sub-micron size range). For the weak irradiation environment ($10^2\, G_0$) the dust traps are located well inside the photoevaporation radius and do not contribute to the dust content in the wind, and for the strong photoevaporative environment ($10^4\, G_0$) there is an increase of $\SI{15}{M_\oplus}$ in the micron range size range for configuration with the inner dust trap respect to the case without dust trap.\par

The dust distribution within the photoevaporative winds should also, in principle, determine the effective cross section and opacity of the disk to the environment FUV irradiation \citep{Facchini2016}. 
If the dust content and effective opacity are high, then the disk could be self-shielded \citep{Qiao2022, Wilhelm2023}, leading to a negative feedback loop where photoevaporation regulates itself. As a first step to study this process we calculate what would be the self-consistent cross section for FUV photons per gas molecule with:
\begin{equation}
\sigma_{FUV} = \sum_a \epsilon_{w}(a)\, \kappa_{UV}(a)\, \mu m_p,
\end{equation}
where $\epsilon_{w}(a)$ is the dust-to-gas ratio of the dust species with size $a$ in the wind, $\kappa_{UV}(a)$ is the absorption opacity at $\lambda = \SI{0.1}{\mu m}$, $\mu= 2.3$ is the mean molecular weight, and $m_p$ is the proton mass \citep[see also][Eq. 23]{Facchini2016}. 
In \autoref{Fig_LostDust_D2G_Sigma_Cum} we show the dust-to-gas ratio in the photoevaporative wind, and the respective FUV cross section per gas molecule (using the Ricci opacities as in the results in Sect.~\ref{sec_Results}).

We find that the total dust-to-gas ratio remains almost constant in the first $\sim$ 0.05-0.1\,Myr of evolution and close to the initial value of $10^{-2}$ for all the irradiation regimes and trap configurations. Once the dust  has had a chance to grow, the dust-to-gas ratio sharply decreases with time as the dust loss rates due to photoevaporation decreases over time (see Fig.~\ref{Fig_LostDust_D2G_Sigma_inst})

Following the same trend, the effective cross section for UV wavelengths is on the order of $\sigma_{\rm{FUV}} \approx 2-\SI{2e-22}{cm^2}$. This is calculated assuming 
\begin{equation}
    \epsilon_{w}(a) = \frac{\dot{\Sigma}_\textrm{d}(a)}{\dot{\Sigma}_\textrm{g}}.
\end{equation}

We also perform the calculation of the FUV cross section assuming an ISM grain size distribution \citep{Mathis1977} for comparison (with $\sigma_{\rm{ISM}} \approx \SI{2.5e-22}{cm^2}$), and find that all our simulations fall below this value, specially after grains have growth after $\sim$0.02-0.1\,Myr.
Though grain growth is expected to reduce the effective absorption cross section at FUV wavelengths \citep{Facchini2016} since larger grains have lower absorption opacities, the decrease of $\sigma_{FUV}$ overtime seems to be mostly dominated by the decrease of the dust-to-gas ratio in the wind.\par

In section \autoref{sec_Discussion_Shielding} we compare the values found for the FUV cross section with previous studies, and discuss if self-shielding by the grains entrained in the wind could be important for the global disk evolution and dispersal process.\par

\section{Discussion}\label{sec_Discussion}

\subsection{Can dust traps explain the observations of highly irradiated regions?} \label{sec_Discussion_Survival}

Observations in the millimeter continuum by \citet{Eisner2018, Otter2021} of the ONC and OMC1 star-forming regions show protoplanetary disks with fluxes on the order of \SI{0.1}{mJy} to \SI{10}{mJy} at \SI{0.85}{}, \SI{1.3}{}, and \SI{3}{mm} wavelengths, as well as a lack of disks sizes larger than \SI{75}{AU} \citep{Otter2021} which could be explained by photoevaporative truncation due to external photoevaporation. \par

However, the work of \citet{Sellek2020} showed that the lifetime of the dust component is significantly shorter in disks undergoing external photoevaporation, than in disks undergoing regular viscous evolution (see \autoref{sec_Resuts_Demo}), with depletion timescales that are on the order of \SI{0.1}{Myrs}. This means that the survival of the dust component in observations could not be explained even in medium radiation environments without an additional process that prevents the dust loss.\par

Since the study of \citet{Sellek2020} does not consider the presence of substructures in protoplanetary disks, which currently are known to be a common feature \citep[e.g. DSHARP sample][]{Andrews2018}, we test whether the presence of dust traps could be help to explain the lifetime of the dust component in photoevaporative disks. 
Though it might seem evident that substructures should contribute to increase the disk lifetime\citep[e.g.][]{Pinilla2012, Pinilla2012_b}, it is not clear whether the dust grains will be able to resist the drag force from the gas in the photoevaporative winds, specially considering that fragmentation continuously replenishes the population of the small grains that are easily entrained with gas.\par

We find in this work that a gap-like substructure can significantly increase the dust component lifetime in weak radiation environments ($10^2\, G_0$) to more than \SI{5}{Myr}, and that in medium radiation environments ($10^3\, G_0$) it can increase the disk lifetime to approx. \SI{2}{Myrs}, but only if the dust trap is located well inside the photoevaporation radius, which in our parameter space was approximately between $50$ and $\SI{75}{AU}$.
The simulations where the dust traps were located outside the photoevaporation radius were dispersed, as the small grains were replenished by fragmentation at a faster rate than the spatial evolution of the photoevaporation front, resulting in the dispersal of the dust traps.\par
For extreme photoevaporation environments ($10^4 G_0$) we observe that substructures have no significant effect on the survival timescale of the protoplanetary disk solid component. The photoevaporation front quickly truncates the whole disk from the outside-in, dragging all the solid grains along with the wind. 
Because this strong irradiation regime is precisely the one that affects the disks in dense stellar regions such as the core of the ONC, we infer that dust traps alone would not be able to explain the millimeter emission in this type of environment.\par

We also consider the possibility that photoevaporation might be delayed, with the environmental FUV flux increasing over time instead of acting at full strength from the very beginning of the simulation. This scenario would be consistent with migration across the stellar cluster \citep{Winter2019}, or shielding of the disk by the primordial envelope \citep{Qiao2022}, however in this case we still find that the material in the dust traps is quickly dispersed once the photoevaporation front reaches the location of the pressure maximum (see \autoref{sec_Results_VariableUV}).

We conclude that the presence of dust traps can increase the disk lifetime in weak and mild photoevaporation environments, which in the case of disks without substructures is limited by the drift timescale at the truncation radius, but cannot help to retain the dust component beyond the lifetime of the gas component, as the grains are efficiently dragged along with the photoevaporative winds. Therefore, while dust traps seem necessary to prevent the dust from draining too quickly by drift, these are not sufficient to explain the survival of the disks in the millimeter continuum, and models that extend simultaneously the lifetime of the gas and dust component such as those accounting for shielding, migration, or multiple star formation events \citep{Qiao2022, Wilhelm2023, Winter2019b_Cyg, Winter2019} are better suited to explain the effective actual disk lifetime.
Other mechanisms that could contribute to the disk long-term survival are the luminosity evolution of bright B stars, which require approximately \SI{1}{Myr} from their formation to reach their peak brightness at UV wavelengths \citep[see][Fig. 3]{Kunitomo2021}, and dominate the irradiation field in regions such as Upper Sco (Anania et al., in prep.), It is necessary to conduct high resolution observations in the millimeter continuum of disks that can be subject to influenced by high environmental irradiation, and identify first if substructures are present, and second if their presence (or absence) would be consistent with the age of the dust component. One promising target for this study would be the disk ISO-Oph2 \citep{Casassus2023}, which is in the proximity of the B star HD147889, and shows signatures of being heated by its bright neighbour.\par

\subsection{Self-shielding by the grains entrained in the wind} \label{sec_Discussion_Shielding}

The calculations performed in this paper use the FRIED grid from \citet{Haworth2018}, which assumes that photoevaporative winds are depleted in dust, with a low dust-to-gas ratio of $\epsilon_w = \SI{3e-4}{}$ and a FUV cross section of $\sigma_{FUV} = \SI{2.7e-23}{cm^2}$.  In Fig.~\ref{Fig_LostDust_D2G_Sigma_inst} we recalculated the FUV cross section using the dust-to-gas ratio measured directly from the simulation lost material, and found them to be higher ($\sigma_{FUV}\approx 10^{-22}$cm$^{-2}$) at early times of evolution ($\sim$0.05\,Myr) than the values used by \citet{Haworth2018}. This means that, in order to be self-consistent with the dust content in the wind, the effective mass loss rates should be lower than those used for this paper \citep[][]{Haworth2023} \par

This self-shielding effect is a negative feedback loop which would moderate the strength of the photoevaporative dispersal. As such we could expect mass loss to proceed at a slower rate, and prolong the disk lifetime in both the gas and dust component, though a dedicated study in which the mass loss and cross section are computed simultaneously in run time would be necessary to confirm this theory.\par

Another important effect to account when accounting for self-shielding would be the spatial evolution of the dust and gas after these are launched from the disk surface (Paine et al., in prep.).  If the ejected material disperses quickly (as suggested in our results in Fig.~\ref{Fig_LostDust_D2G_Sigma_inst}), it will not be able to shield the disk from the external irradiation sources since the cross section sharply drops after the first few $\sim$0.02-0.1\,Myr of disk evolution.
Therefore, in order to determine if self-shielding can extend the disk lifetime or not, it is also necessary to study whether the dust and gas material lifted by the photoevaporative winds can remain around their parent protoplanetary disk for longer timescales, on the scale of $\SI{1}{Myr}$ or more.\par

We note that in this work we assumed the opacity values of \citet{Ricci2010} to calculate the FUV cross sections, while previous calculations of \citet[][for example]{Facchini2016}, used the optical constants from \citet{Li1997} which leads to the difference in the reported for the ISM cross section, where our estimation is lower approximately by a factor of $4$. We infer that our cross sections would increase by a factor of a few if we had used the same optical constants and grain composition.\par

Previous studies have shown that shielding can reduce the effect of external photoevaporation on the disk \citep[e.g.][]{Qiao2022}, however its effect is usually constrained to the early stages of disk evolution. 
In contrast, if the self-shielding by the wind-entrained grains is proven to be effective, it could actually contribute to reduce the mass loss in high irradiation environments for extended periods of time, specially in disks with multiple substructures.\par

We highlight the importance of developing self-consistent models that couple dust trapping, photoevaporation driven by external irradiation, and the dust content in the ejected winds, since these ingredients do interact with each other and change the course of the disk evolution, as shown, for example, in \citet{Owen2023} where it was proposed that circumstellar disks in extreme environments such as the galactic centre could survive photoevaporative dispersal.

\section{Summary}\label{sec_Summary}

In this work, we studied the evolution of the dust component and emission in the millimeter continuum of protoplanetary disks subject to high environmental FUV irradiation, accounting for the presence of gap-like substructures that act as dust traps.
As in \citet{Sellek2020}, we also find that dust drift is more efficient in disks subject to external photoevaporation than in standard viscous disks, where the lifetime of the dust component can be as short as a few \SI{0.1}{Myr} if dust traps are not present.\par
In weak irradiation environments of $F_{UV} = 10^2\, G_0$, the presence of dust traps does prevent drift of dust grains into the star and allow the disk to survive for several Myrs. 
In irradiation environments of $10^3\, G_0$, dust traps need to be inside the photoevaporative truncation radius to extend the disk lifetime (from \SI{0.3}{} to \SI{2.3}{Myr} for our choice of parameters, see \autoref{sec_Results_ParamSpace}), while dust traps located outside the truncation radius are dispersed with the photoevaporative winds and do not extend the lifetime of the dust component significantly. 
Finally, in extreme irradiation environments of $10^4\, G_0$ all the dust traps are dispersed as the photoevaporation front clears the entire disk from the outside-in.\par

Though dust traps are a necessary ingredient to explain the survival of the dust and the observed millimeter emissions in highly irradiated environments, specially considering that drift is specially efficient in disks truncated by external photoevaporation \citep{Sellek2020}, they are not enough to explain why the objects found at the core of dense stellar regions such as the ONC or Cyg OB2 \citep{Guarcello2016, Eisner2018, Otter2021} have not yet dispersed.

Overall, it seems more likely that the disks observed in these regions have only recently began to experience the high irradiation and extreme mass loss rates measured in observations. 
These objects could have been subject to a much lower irradiation earlier in their evolution due to shielding, migration from the regions with lower stellar densities, or they could have been born in a more recent star formation event than the rest of the cluster \citep{Winter2019b_Cyg, Winter2019, Qiao2022}. \par

On that line, we do find that the dust content entrained with the photoevaporative winds might be enough to increase the cross section for FUV photons to $\sigma_{FUV} \approx \SI{e-22}{cm^2}$ at \SI{0.1}{\mu m} wavelengths at early times of evolution, partially shielding the protoplanetary disk from external irradiation, and decreasing the total mass loss rate. However, it is yet unclear whether the dusty material lifted with the winds will stay in the proximity of the parent disk, effectively acting as a shield, or if it will quickly disperse leaving the disk surface bare to the environmental irradiation, and further studies are necessary.\par

In conclusion, dust traps are necessary but not sufficient to explain the survival of the dust component and continuum emission in high photoevaporative regions, and self-shielding by the dust entrained with the wind may contribute to reduce the extreme mass loss rates, but only if the dusty material can remain close enough to the parent disk to absorb the irradiation from the environment.

\begin{acknowledgements}
MG and PP acknowledge funding from the Alexander von Humboldt Foundation in the framework of the Sofja Kovalevskaja Award endowed by the Federal Ministry of Education and Research. PP acknowledge funding from UKRI under the UK government’s Horizon Europe funding guarantee from ERC. TJH is funded by a Royal Society Dorothy Hodgkin Fellowship and UKRI guarantee funding (EP/Y024710/1). SF is funded by the European Union under the European Union's Horizon Europe Research \& Innovation Programme 101076613 (UNVEIL). Views and opinions expressed are however those of the author(s) only and do not necessarily reflect those of the European Union or the European Research Council. Neither the European Union nor the granting authority can be held responsible for them.

\end{acknowledgements}
\bibpunct{(}{)}{;}{a}{}{,} 
\bibliographystyle{aa} 
\bibliography{TheBibliography.bib} 

\end{document}